\documentclass[a4paper,11pt]{article}
\pdfoutput=1 

\usepackage{jcappub} 

\usepackage[T1]{fontenc} 
\usepackage{subcaption} 

\title{\boldmath Modified Cosmology from Mass-to-Horizon Relation: Observational Bounds}

\author{Pranav Prasanthan,}
\author{Hussain Gohar,}
\author{and Vincenzo Salzano}

\affiliation{Institute of Physics, University of Szczecin, Wielkopolska 15, 70-451 Szczecin, Poland}

\emailAdd{pranav.prasanthan@phd.usz.edu.pl}
\emailAdd{hussain.gohar@usz.edu.pl}
\emailAdd{vincenzo.salzano@usz.edu.pl}

\abstract{We constrain the class of modified cosmologies derived in Paper I \cite{paper-I} from a
generalized mass-to-horizon relation (MHR) that enforces thermodynamic
consistency between the Cai--Kim horizon temperature and generalized
horizon entropies. The modified Friedmann equations depend on an
entropy exponent $m$, an MHR coupling parameter $\gamma$, and an
entanglement-correction amplitude $f_B$, with standard $\Lambda$CDM
recovered in the appropriate limit. Using Pantheon$+$/SH0ES
Type~Ia supernovae, cosmic chronometers, DESI DR2 baryon acoustic
oscillations, and Planck 2018 CMB distance priors, we constrain
eight physically motivated sub-cases via Markov chain Monte Carlo
and compare models through the Bayesian log-evidence.
The entropy exponent is
tightly bounded, $|m-1|\lesssim\mathcal{O}(10^{-4})$ when the MHR
coupling is fixed ($\gamma=1$), relaxing to $\mathcal{O}(10^{-3})$
along the $m$--$\gamma$ degeneracy, excluding
any macroscopically significant departure from standard horizon
thermodynamics. Freeing the MHR coupling parameter or the
entanglement amplitude raises the
inferred Hubble constant to $h\simeq0.70$--$0.71$, reducing the
CMB--SH0ES tension from ${\sim}4\sigma$ to
${\sim}1.2$--$2.6\sigma$, but no scenario fully resolves it
within a flat universe. The Bayesian log-evidence nevertheless
disfavors every extension relative to $\Lambda$CDM in all
dataset combinations
($-16\lesssim\Delta\ln\mathcal{Z}\lesssim-1$): the improved fits
obtained when the SH0ES calibration is included reflect an
absorption of the Hubble tension by the additional parameters
rather than genuine evidence for modified horizon entropy.}

\begin{document}
\maketitle
\flushbottom

\section{Introduction}
\label{sec:introduction}

The thermodynamic nature of gravitational horizons, pioneered by
Bekenstein and Hawking~\cite{Bekenstein:1973ur,Hawking:1974rv}, has
profoundly influenced our understanding of black holes and cosmology.
The Bekenstein--Hawking entropy,
$S_{\text{BH}} = k_B c^3 A / (4\hbar G)$, and the associated Hawking
temperature, $T_{\text{H}} = \hbar \kappa / (2\pi k_B c)$, lay the
foundation for the gravity--thermodynamics
correspondence~\cite{Jacobson:1995ab,Cai:2005ra,Padmanabhan:2003pk,
Verlinde:2010hp}.  This framework, particularly the first law of
thermodynamics applied to the apparent horizon of a
Friedmann--Lema\^itre--Robertson--Walker (FLRW) universe, has proven
instrumental in deriving the Friedmann equations and exploring
modifications thereof~\cite{Cai:2005ra,Gong:2007md}.  A crucial,
yet often implicit, assumption in such applications is the existence
of a mass-to-horizon relation (MHR)~\cite{Gohar:2023hnb,
Gohar:2023lta,Gohar:2025yfx,Gohar:2026hiy, Denkiewicz:2025txx}, which links the mass (or energy)
enclosed by the horizon to its size.

The simplest linear MHR, $M = \gamma (c^2/G) L$, is a natural
extrapolation from the Schwarzschild black hole case.  However, a
growing body of literature has explored extensions of the
Bekenstein--Hawking entropy motivated by non-extensive
statistics~\cite{Tsallis:1987eu,Tsallis:2012js,reny1}, quantum
gravity effects~\cite{Rovelli:1996dv,Meissner:2004ju,Medved:2004yu},
fractal structure of the horizon~\cite{Barrow:2020tzx}, and the
entanglement entropy of quantum fields across a
horizon~\cite{Das:2007mj}.  These generalized entropies, when
combined with the Hawking temperature, often lead to thermodynamic
inconsistencies~\cite{Cimdiker:2022ics} in holographic scenarios if
the underlying linear MHR is not also
generalized~\cite{Gohar:2023lta,Gohar:2025yfx}.

A unified theoretical framework that addresses these issues was
recently formulated by two of the authors of this
article~\cite{Gohar:2023lta} and subsequently generalized
in~\cite{Gohar:2025yfx} to a broader class of models, encompassing
most of the known generalizations of the Bekenstein entropy.
Following~\cite{Gohar:2025yfx, Gohar:2026hiy}, the generalized MHR takes the form
\begin{equation}
M = \gamma \frac{c^2}{G} \ell_{\text{Pl}} \left[ \frac{L}{\ell_{\text{Pl}}}
  \mp \beta \left( \frac{L}{\ell_{\text{Pl}}} \right)^{3-\alpha}
  \right]^m,
\label{eq:MHR}
\end{equation}
where $L$ denotes the cosmological horizon radius (identified in
what follows with the apparent-horizon radius $r_a$),
$\ell_{\text{Pl}}=\sqrt{\hbar G/c^3}$ is the Planck length,
$\gamma>0$ and $\beta>0$ are dimensionless parameters, $m>0$ and
$\alpha \in \mathbb{R}$ are real exponents, and the upper (lower)
sign encodes a quantum/entanglement correction term.  This relation
ensures thermodynamic consistency between the horizon temperature
and a broad class of generalized entropies; throughout this work
we adopt the Cai--Kim temperature,
$T_h = \hbar c/(2\pi k_B r_a)$~\cite{Cai:2005ra}, the
positive-definite temperature associated with the apparent
horizon of a FLRW universe.  By invoking the Clausius
relation $dE = T\,dS$, one obtains the corresponding generalized
mass-to-horizon entropy~\cite{Gohar:2025yfx, Gohar:2026hiy}
\begin{equation}
S_{\text{G}} = 2\pi \gamma k_B \left[
  \frac{m}{m+1}\left(\frac{r_a}{\ell_{\text{Pl}}}\right)^{m+1}
  \mp m\beta \frac{\sigma-1}{\sigma}
  \left(\frac{r_a}{\ell_{\text{Pl}}}\right)^{\sigma} \right],
\label{eq:entropy_def}
\end{equation}
with $\sigma = m + 3 - \alpha$. In appropriate limits and upon a suitable reparametrization of \(m\), \(S_G\) reduces to the Tsallis--Cirto ($m=2\delta-1$)~\cite{Tsallis:1987eu,Tsallis:2012js} and Barrow-type ($m=1+\Delta$)~\cite{Barrow:2020tzx} entropies, as well as to quantum-gravity- and entanglement-induced corrections to both the Bekenstein and Tsallis--Cirto and Barrow-type entropies. Interestingly, the MHR mass and \(S_G\) are not merely ad hoc generalizations introduced to ensure holographic thermodynamic consistency; rather, they establish a direct connection between entropy functionals in statistical mechanics and scalar–tensor theories of gravity via the Misner–Sharp mass and the Wald entropy functional \cite{Gohar:2026hiy}.

In Paper I~\cite{paper-I} (see also the discussion in
Sec.~\ref{sec:modified} below), we have employed the
gravity--thermodynamics correspondence---specifically, the first law of thermodynamics
applied on the apparent horizon with the Cai--Kim temperature---to
this generalized entropy, thereby deriving the modified Friedmann
equations that govern the cosmological dynamics.  The standard
$\Lambda$CDM scenario is recovered for $m = \gamma = 1$ and
$\beta = 0$.

The present work is devoted to a comprehensive observational test of
this modified cosmology.  We utilize a combination of the latest
geometric datasets: the Pantheon+ compilation of Type~Ia supernovae
(including the SH0ES Cepheid calibration)~\cite{Brout:2022vxf},
cosmic chronometers measurements of $H(z)$~\cite{Jiao:2022aep},
baryon acoustic oscillations from DESI DR2~\cite{DESI:2025zgx}, and
Planck 2018 CMB distance priors~\cite{Planck:2018vyg}.  The analysis
is carried out with a Markov chain Monte Carlo (MCMC) method using
the \texttt{emcee} sampler~\cite{Foreman-Mackey:2012any}.  Model
comparison is performed via the Bayesian log-evidence computed
from the chains.  The modified Hubble
expansion $H(z)$ is computed by numerically solving the implicit
Friedmann equation derived from Eq.~\eqref{eq:entropy_def}.  To
explore the model parameter space meaningfully, we consider several
physical sub-cases by fixing a subset of parameters
(Table~\ref{tab:cases}).  For models with correlated parameters
($m$--$\gamma$ and $\alpha$), physically motivated
reparametrizations are employed to remove known degeneracies and
improve sampler efficiency (Sec.~\ref{sec:reparam}).

The paper is structured as follows.  In Sec.~\ref{sec:modified} we
summarize the essential equations of the modified cosmology,
referring to Paper I \cite{paper-I} for the full thermodynamic
derivation.  Section~\ref{sec:data} describes the observational
datasets and the statistical methodology.  Our results are presented
and discussed in Sec.~\ref{sec:results}, organized thematically:
parameter constraints for the area-law (Case I) and entanglement
(Case II) families of Paper I, the response of the Hubble tension
across scenarios, Bayesian model comparison, and a comparison with
previous studies.  Our conclusions are
given in Sec.~\ref{sec:conclusion}.

\section{Modified Cosmology from Mass-To-Horizon Entropy}
\label{sec:modified}

The full thermodynamic derivation of the modified Friedmann
equations using the first law of thermodynamics on the apparent
horizon and generalized mass-to-horizon entropy is provided in the
companion theoretical paper, Ref.~\cite{paper-I}.  Here we
summarize only the essential equations required for the data
analysis; the present paper is self-contained at the level of the
equations used in the MCMC analysis.

Inserting the entropy~\eqref{eq:entropy_def} into the first law,
together with the Cai--Kim temperature $T_h$
(Sec.~\ref{sec:introduction}) and the energy flux
$\delta Q = -dE = 4\pi r_a^3 H(\varepsilon+P) \,dt$, yields the
modified Friedmann equation for a spatially flat universe ($k=0$):
\begin{equation}\label{eq:friedmann_flat}
2m\gamma c^2\left[\frac{1}{3-m}
\frac{\left(\frac{H}{c}\right)^{3-m}}{\ell_{\rm Pl}^{m-1}}
\mp\beta\frac{\sigma-1}{4-\sigma}
\frac{\left(\frac{H}{c}\right)^{4-\sigma}}
{\ell_{\rm Pl}^{\sigma-2}}\right]
= \frac{8\pi G}{3c^2}\varepsilon + \frac{\Lambda c^2}{3},
\end{equation}
where $\varepsilon = \varepsilon_m + \varepsilon_r$ is the total
energy density of matter and radiation, and $\Lambda$ is the
cosmological constant, which arises as an integration constant of
the first law and is retained as in Paper I~\cite{paper-I}.  This
equation can be recast in the standard form
\begin{equation}\label{eq:dimensionless_friedmann}
H^2 = \frac{8\pi G}{3c^2}\left(\varepsilon_m + \varepsilon_r
  + \varepsilon_{de}\right),
\end{equation}
with an effective dark energy density
\begin{equation}\label{eq:ede}
\varepsilon_{de} = \frac{3c^2}{8\pi G}\left[
\frac{\Lambda c^2}{3} + H^2\left(1
- \frac{2m\gamma}{(3-m)\ell_{\rm Pl}^{m-1}}
\left(\frac{H}{c}\right)^{1-m}\right.\right.
\left.\left.\pm\frac{2m\gamma\beta(\sigma-1)}
{(4-\sigma)\ell_{\rm Pl}^{\sigma-2}}
\left(\frac{H}{c}\right)^{2-\sigma}\right)\right].
\end{equation}
We introduce the normalized Hubble parameter
$E(a)\equiv H(a)/H_0$, where $a=(1+z)^{-1}$ is the scale factor.
Defining the present-day density parameters $\Omega_{m0}$ and
$\Omega_{r0}$ for matter and radiation, the modified Friedmann
equation becomes implicit:
\begin{equation}\label{eq:Ea}
E^2(a) = \frac{\Omega_{m0} a^{-3} + \Omega_{r0} a^{-4}}
              {1- \Omega_{de}(a,E)},
\end{equation}
with
\begin{equation}\label{eq:Omega_deEA}
\Omega_{de}(a,E) = \frac{\Omega_{\Lambda 0}}{E^2(a)}
  + 1- A\,E^{1-m} \pm B\,E^{2-\sigma},
\end{equation}
where $\Omega_{\Lambda 0}\equiv\Lambda c^2/(3H_0^2)$ is the
density parameter associated with the cosmological constant,
and the dimensionless parameters
\begin{equation}\label{eq:AB}
A = \frac{2m\gamma}{3-m}\left(\ell_{\rm Pl}\frac{H_0}{c}\right)^{1-m},\
B = \frac{2m\gamma\beta(\sigma-1)}{4-\sigma}
    \left(\ell_{\rm Pl}\frac{H_0}{c}\right)^{2-\sigma}.
\end{equation}
The constant $\Omega_{\Lambda 0}$ is fixed by the present-day
flatness condition $1=\Omega_{m0}+\Omega_{r0}+\Omega_{de0}$,
yielding
\begin{equation}\label{eq:OmLambda}
\Omega_{\Lambda 0} = A \mp B - (\Omega_{m0}+\Omega_{r0}).
\end{equation}
\emph{Sign convention.}\ Equations~\eqref{eq:MHR}--\eqref{eq:OmLambda}
are to be read consistently with the upper (or lower) signs
throughout: the upper branch of the MHR~\eqref{eq:MHR} corresponds
to the upper sign in Eqs.~\eqref{eq:Omega_deEA}
and~\eqref{eq:OmLambda}, for which the $B$-term contributes
positively to $\Omega_{de}$.  All results in this paper are
obtained on this branch (Sec.~\ref{sec:cases}), which is also the
branch analyzed in Paper I~\cite{paper-I}.

We stress that the effective density $\varepsilon_{de}$ is a
bookkeeping device rather than an independent fluid: it depends
explicitly on $H$ and therefore on the total energy content of
the universe.  Its effective equation of state $\omega_{de}(a)$,
defined and analyzed in Paper I~\cite{paper-I}, tracks the
dominant background component at early times and relaxes to
$\omega_{de}=-1$ today, with a tracking amplitude set by the
deviations of $A$ (equivalently $\gamma$) from unity and of the
entanglement amplitude from zero.  Given the tight observational
bounds on these deviations derived below, we do not analyze
$\omega_{de}(a)$ further here and refer to Paper I for its
behavior across the parameter space.


Equation~\eqref{eq:friedmann_flat} is undefined at $m=3$, where
the prefactor $(3-m)^{-1}$ diverges, and at $\sigma=4$, where
$(4-\sigma)^{-1}$ diverges.  These are not merely technical
singularities but signal genuine pathologies in the thermodynamic
derivation: as $m\to 3$, the leading entropy term scales as
$r_a^4$ (rather than $r_a^2$), so the area--entropy correspondence
breaks down.  As $\sigma\to 4$, the subleading correction term
grows faster than the leading term at large $r_a$, destabilizing
the thermodynamic equilibrium of the horizon.  In the entanglement
scenarios the condition $\sigma=4$ corresponds to $\alpha=m-1$;
for $m=1$ this is $\alpha=0$, i.e.\ a correction that scales as
$r_a^4$, consistent with the $m=3$ pathology.

A second potential pathology arises from the denominator
$(1-\Omega_{de})$ in Eq.~\eqref{eq:Ea}, which vanishes when the
effective dark energy density equals the critical density.  For
parameter values near the singular limits, this can occur at finite
redshift, producing a cosmological singularity or a bounce.  The
two pathologies are handled differently in the analysis.  The
singular limits are excluded directly by the priors: the prior
$m\in(0.5,2.5)$ keeps the sampler well away from $m=3$, and
proposals with $|\sigma-4|<0.1$ are rejected.  The second
pathology cannot be excluded by fixed prior bounds, because the
redshift at which $1-\Omega_{de}$ may vanish depends on the full
parameter combination; it is instead handled at run time: any
proposal yielding $E^2\leq 0$ or $1-\Omega_{de}\leq 0$ anywhere
in the integration range is rejected by assigning $-\infty$ to
the log-likelihood.

\section{Observational Data and Statistical Methodology}
\label{sec:data}

This section describes the cosmological datasets employed to
constrain the MHR modified cosmology summarized in
Sec.~\ref{sec:modified}.  The analysis uses the latest available
geometrical observations: Type~Ia supernovae (SNeIa) from
Pantheon$+$ (with SH0ES anchors), cosmic chronometers (CC)
measurements of $H(z)$, baryon acoustic oscillations (BAO) from
DESI-DR2, and Planck 2018 cosmic microwave background (CMB) distance
priors.  After presenting each dataset, we define the physical
model cases, the reparametrizations used by the sampler, and the
MCMC implementation.

The SNeIa sample consists of distance moduli derived from 1701
light curves corresponding to 1550 spectroscopically confirmed
events, taken from the Pantheon$+$ compilation~\cite{Brout:2022vxf}.
The redshift coverage extends from $0.001<z<2.26$.  The
corresponding chi-squared statistic is defined as
\[
\chi^2_{SN} = \Delta \boldsymbol{\mathcal{\mu}}^{SN}
  \cdot \mathbf{C}^{-1}_{SN}
  \cdot \Delta \boldsymbol{\mathcal{\mu}}^{SN},
\]
where $\Delta\boldsymbol{\mathcal{\mu}} =
\mathcal{\mu}_{\rm theo} - \mathcal{\mu}_{\rm obs}$ represents the
difference between theoretical and observed distance moduli, and
$\mathbf{C}_{SN}$ denotes the total (statistical plus systematic)
covariance matrix.  The theoretical distance modulus is computed from
\[
\mu_{\rm theo}(z_{hel},z_{HD},\boldsymbol{p}) = 25
  + 5 \log_{10} \bigl[ d_{L}(z_{hel}, z_{HD}, \boldsymbol{p}) \bigr],
\]
with the luminosity distance $d_L$ (in Mpc) given by
\[
d_L(z_{hel}, z_{HD},\boldsymbol{p})
  = (1+z_{hel})\int_{0}^{z_{HD}}
    \frac{c\,dz'}{H(z',\boldsymbol{p})}.
\]
Here $z_{hel}$ is the heliocentric redshift, $z_{HD}$ the
Hubble-diagram redshift~\cite{Carr:2021lcj}, and $\boldsymbol{p}$
the vector of cosmological parameters.  The observed distance
modulus is $\mu_{obs} = m_B - M$, where $m_B$ is the standardized
blue apparent magnitude and $M$ the fiducial absolute magnitude
calibrated via primary distance anchors.

Within the Pantheon$+$ sample, 77 SNeIa reside in Cepheid-host
galaxies whose distance moduli $\mu_{Ceph}$ are independently
calibrated, thereby breaking the usual degeneracy between $H_0$ and
$M$.  Consequently, the residual vector is constructed as
\[
\Delta\boldsymbol{\mathcal{\mu}} =
\begin{cases}
m_{B,i} - M - \mu_{Ceph,i}, & i \in \text{Cepheid hosts},\\[4pt]
m_{B,i} - M - \mu_{{\rm theo},i}, & \text{otherwise}.
\end{cases}
\]
We refer to this combined dataset as PP\&SH0ES.

The Hubble parameter $H(z)$ is measured using the cosmic chronometer
approach (CC)~\cite{Jimenez:2001gg,Moresco:2010wh,Moresco:2018xdr,
Moresco:2020fbm,Moresco:2022phi}, which relies on early-type
galaxies undergoing passive evolution.  The most recent compilation,
presented in Ref.~\cite{Jiao:2022aep}, spans $0<z<1.965$ and
contains 33 measurements.  The corresponding chi-squared is
\[
\chi^2_{H} = \Delta \boldsymbol{\mathcal{H}}
  \cdot \mathbf{C}^{-1}_{H}
  \cdot \Delta \boldsymbol{\mathcal{H}},
\]
where $\Delta \boldsymbol{\mathcal{H}} = H_{\rm theo} - H_{\rm data}$
and $\mathbf{C}_{H}$ is the total covariance matrix computed
following Ref.~\cite{Moresco:2020fbm}.

CMB constraints are implemented via a compressed likelihood approach
based on the shift parameters~\cite{Wang:2007mza,Zhai:2019nad}.  In
particular, we employ the quantities
\[
R(\boldsymbol{p}) \equiv
  \frac{\sqrt{\Omega_m H^2_{0}}\, r(z_{\ast},\boldsymbol{p})}{c},
\qquad
l_{a}(\boldsymbol{p}) \equiv
  \frac{\pi\, r(z_{\ast},\boldsymbol{p})}
       {r_{s}(z_{\ast},\boldsymbol{p})},
\]
updated to the Planck 2018 data release~\cite{Planck:2018vyg}.  The
chi-square is
\[
\chi^2_{CMB} = \Delta \boldsymbol{\mathcal{F}}^{CMB}
  \cdot \mathbf{C}^{-1}_{CMB}
  \cdot \Delta \boldsymbol{\mathcal{F}}^{CMB},
\]
where the vector $\boldsymbol{\mathcal{F}}^{CMB}$ includes $R$,
$l_a$, the baryon density $\Omega_b h^2$, and the dark-matter
density $(\Omega_{m,0}-\Omega_{b,0})h^2$.  The photon-decoupling
redshift $z_{\ast}$ is computed using the fitting formula of
Ref.~\cite{Aizpuru:2021vhd}.  The comoving distance at decoupling
is $r(z_{\ast},\boldsymbol{p}) = \int_0^{z_{\ast}}
c\,dz'/H(z',\boldsymbol{p})$, and the comoving sound horizon is
\[
r_s(z,\boldsymbol{p}) = \int_z^{\infty}
  \frac{c_s(z')}{H(z',\boldsymbol{p})}\, dz',
\]
with sound speed $c_s(z) = c/\sqrt{3(1+\overline{R}_b (1+z)^{-1})}$
and baryon-to-photon density ratio $\overline{R}_b =
31500\,\Omega_b h^2\,(T_{\rm CMB}/2.7)^{-4}$, taking
$T_{\rm CMB}=2.726$~K.

We note an important caveat concerning the CMB compressed
likelihood: the shift-parameter compression has been
validated primarily for $w$CDM-type dark energy
models~\cite{Wang:2007mza,Zhai:2019nad}; its accuracy for the MHR
scenarios, where the effective dark energy density depends
explicitly on $H$ (see Eq.~\eqref{eq:ede}), has not been
independently established.  In particular, the fitting formula for
$z_*$ from Ref.~\cite{Aizpuru:2021vhd} was calibrated for smooth
dark energy models, and modest systematic errors in its evaluation
for the MHR scenarios cannot be excluded.  This approximation
should be revisited in a full CMB power-spectrum analysis.

BAO measurements are taken from the latest data release of the Dark
Energy Spectroscopic Instrument
(DESI-DR2)~\cite{DESI:2025zgx}.  The sample includes Bright
Galaxies, Luminous Red Galaxies, Emission Line Galaxies, Quasars,
and Lyman-$\alpha$ forest tracers, covering redshifts
$0.1 \lesssim z \lesssim 4.2$ (see Table~IV of
Ref.~\cite{DESI:2025zgx}) and comprising 13 data points.  The
observables are the ratios
$D_M(z,\boldsymbol{p})/r_s(z_d,\boldsymbol{p})$,
$D_H(z,\boldsymbol{p})/r_s(z_d,\boldsymbol{p})$, and
$D_V(z,\boldsymbol{p})/r_s(z_d,\boldsymbol{p})$, where
$D_M(z) = \int_0^{z} c\,dz'/H(z')$ is the transverse comoving
distance (coinciding with the line-of-sight comoving distance in
a spatially flat universe~\cite{Hogg:1999ad}),
$D_H(z) = c/H(z)$ is the Hubble distance, and $D_V$ is the
angle-averaged distance defined below, all normalized by the
comoving sound horizon at the drag epoch, $r_s(z_d)$.  The sound
horizon is evaluated from the same integral as $r_s(z_*)$, with
the drag-epoch redshift $z_d$ computed using the fitting formula
of Ref.~\cite{Aizpuru:2021vhd}, as for $z_*$.  Correlations
between $D_M/r_s$ and $D_H/r_s$ are fully accounted for in the
covariance matrix.  The chi-square is
\[
\chi^2_{BAO} = \Delta \boldsymbol{\mathcal{F}}^{BAO}
  \cdot \mathbf{C}^{-1}_{BAO}
  \cdot \Delta \boldsymbol{\mathcal{F}}^{BAO},
\]
with $\boldsymbol{\mathcal{F}}^{BAO}$ comprising the three distance
ratios.  The volume-averaged distance is defined as
\[
D_V(z,\boldsymbol{p}) = \left[ c\,z\,(1+z)^2\,
  D_A^2(z,\boldsymbol{p}) / H(z,\boldsymbol{p}) \right]^{1/3},
\]
and the angular diameter distance $D_A = D_M/(1+z)$.

\subsection{Model cases and free parameters}
\label{sec:cases}

The extended MHR entropy~\eqref{eq:entropy_def} contains four
continuous parameters ($m,\gamma,\alpha,\beta$) and a discrete sign
choice; we restrict to the upper MHR branch
(Sec.~\ref{sec:modified}).  Different
physical regimes are selected by fixing subsets of these parameters,
as summarized in Table~\ref{tab:cases}.  The naming follows the
entropy families that each case generalizes---BH
(Bekenstein--Hawking, $m=1$), BTC (Barrow--Tsallis--Cirto, free
$m$), and ENT (entanglement-corrected, $\beta=1$)---with the Roman
numeral incremented when the MHR coupling $\gamma$ is also freed.
In the classification of Paper I~\cite{paper-I}, the BH and BTC
scenarios realize the standard area-law extensions (Case~I,
$\beta=0$) and the ENT scenarios its quantum-entanglement
corrections (Case~II, $\beta=1$); the quantum-gravity corrections
(Case~III, $\alpha=4$) are not analyzed separately, since their
amplitude $B\propto(\ell_{\rm Pl}H_0/c)^{3-m}\sim10^{-122}$
renders them observationally indistinguishable from the
corresponding Case~I scenario.  The
baseline cosmological parameters $\Omega_{m0}$, $\Omega_{b0}$, $h$
are always sampled.

\begin{table}[h]
\centering
\caption{Modified cosmological scenarios examined in the MCMC
  analysis.  The standard $\Lambda$CDM limit is recovered for
  $m=1$, $\gamma=1$, and $\beta=0$.  Baseline parameters
  $\{\Omega_m, \Omega_b, h\}$ are always sampled.  The ``Free''
  column lists the physically free model parameters; the ``Sampled''
  column lists the additional parameters that
  are passed directly to the \texttt{emcee} sampler, with derived
  parameters recovered analytically after each step
  (Sec.~\ref{sec:reparam}).  The ``Case'' column gives the
  corresponding classification of Paper I~\cite{paper-I}.
  All cases use the upper MHR sign branch.}
\label{tab:cases}
\begin{tabular}{lcccl}
\multicolumn{1}{c}{\bf Model} &
\multicolumn{1}{c}{\bf Case} &
\multicolumn{1}{c}{\bf Free} &
\multicolumn{1}{c}{\bf Sampled} &
\multicolumn{1}{c}{\bf Fixed} \\
\hline
BH-I   & I  & --                     & --                   & $m=1$, $\gamma=1$, $\beta=0$ \\
BH-II  & I  & $\gamma$               & $A$                  & $m=1$, $\beta=0$ \\
BTC-I  & I  & $m$                    & $m$                  & $\gamma=1$, $\beta=0$ \\
BTC-II & I  & $m$, $\gamma$          & $A$, $m$             & $\beta=0$ \\
BH ENT-I  & II & $\alpha$            & $f_B$                & $m=1$, $\gamma=1$, $\beta=1$ \\
BH ENT-II & II & $\alpha$, $\gamma$  & $A$, $f_B$           & $m=1$, $\beta=1$ \\
BTC ENT-I & II & $m$, $\alpha$       & $m$, $f_B$           & $\gamma=1$, $\beta=1$ \\
BTC ENT-II& II & $m$, $\alpha$, $\gamma$ & $A$, $m$, $f_B$ & $\beta=1$ \\
\end{tabular}
\end{table}

\subsection{Physically motivated reparametrizations}
\label{sec:reparam}

Several model cases suffer from strong parameter degeneracies that
degrade sampler efficiency; we break them by sampling
reparametrized combinations and recovering the physical parameters
analytically after each step.

\paragraph{Auxiliary scale ratio.}
All modified terms depend on the single dimensionless Planck-to-Hubble
ratio
\begin{equation}\label{eq:X}
  X \equiv \frac{\ell_{\rm Pl}\,H_0}{c}
  \,=\,\frac{\ell_{\rm Pl}}{c}\,100\,h\,\frac{\mathrm{km\,s}^{-1}}
  {\mathrm{Mpc}}\;\approx\;1.75\times10^{-61}\,h,
\end{equation}
with $\ell_{\rm Pl}=\sqrt{G\hbar/c^3}\approx1.616\times10^{-35}$~m.
In terms of $X$, Eqs.~\eqref{eq:AB} read
$A=(2m\gamma/(3-m))\,X^{1-m}$ and
$B=(2m\gamma\beta(\sigma-1)/(4-\sigma))\,X^{2-\sigma}$.

\paragraph{BH-II ($m=1$ fixed, $\gamma$ free).}
At $m=1$ the exponent $X^{1-m}=1$, so $A=\gamma$ exactly; we
sample $A$. 

\paragraph{BTC-II ($m$ and $\gamma$ both free).}
Near $m=1$ the combination $A=(2m\gamma/(3-m))\,X^{1-m}$ is
the only combination of $m$ and $\gamma$ that controls
$\Omega_{\Lambda0}=A-(\Omega_{m0}+\Omega_{r0})$.  A joint shift
$(\delta m,\,\delta\gamma)$ that preserves $A$ satisfies
\begin{equation}\label{eq:degeneracy_condition}
  \frac{\delta\gamma}{\gamma}
  \approx -\ln\!\left(\frac{c}{\ell_{\rm Pl} H_0}\right)\delta m
  \approx -140\,\delta m,
\end{equation}
producing a near-flat ridge in the $(m,\gamma)$ plane with slope
$\approx-140$ that direct sampling handles poorly.  We break the
degeneracy by sampling $(A,\,m)$ instead of $(\gamma,\,m)$; the
physical parameter is recovered after each step via
\begin{equation}\label{eq:gamma_from_A}
  \gamma = \frac{A\,(3-m)}{2m}\,X^{m-1}.
\end{equation}
To preserve a flat prior on $(\gamma,m)$ when sampling $(A,m)$, the
log-posterior is corrected by the log-Jacobian
\begin{align}\label{eq:log_jac}
  \ln\left|\frac{\partial\gamma}{\partial A}\right|_m
  &= \ln\!\frac{3-m}{2m} + (m-1)\ln X \notag\\
  &= \ln(3-m) - \ln(2m) + (m-1)\ln X.
\end{align}
This correction is large for $m\neq1$ ($\ln X\approx-140$) and
cannot be neglected.

\paragraph{BH ENT-I and related entanglement models ($\alpha$ free).}
Near $m=1$ the $B$-term scales as $X^{\alpha-2}$, so a shift
$\delta\alpha$ rescales $B$ by
$X^{\delta\alpha}\approx10^{-61\,\delta\alpha}$: the mapping from
$\alpha$ to the likelihood is exponentially steep and poorly
conditioned.  We therefore replace
$\alpha$ by the fractional correction
\begin{equation}\label{eq:fB}
  f_B \;\equiv\; \frac{B}{(3-m)\,A}
  \;=\;\frac{\beta(\sigma-1)}{4-\sigma}\,X^{\alpha-2},
\end{equation}
which is well-conditioned and approximately linear over the prior
range; all entanglement cases fix $\beta=1$.  The physical
$\alpha$ is recovered after each step by Newton iteration on the
log-transformed residual
\begin{equation}\label{eq:Newton_h}
  h(\alpha) = \ln\!\frac{\sigma-1}{4-\sigma}
             + (\alpha-2)\ln X - \ln f_B = 0,
\end{equation}
with $\sigma=m+3-\alpha$.  No additional Jacobian correction is
required because the prior is stated directly on $f_B$ (uniform
on $[0,f_{B,\rm max}]$).

\paragraph{BTC ENT-II (most general).}
The sampled vector is $(A,\,m,\,f_B)$.  After each step,
$\gamma$ is recovered from Eq.~\eqref{eq:gamma_from_A} and $\alpha$
from Eq.~\eqref{eq:fB}, and the BTC-II Jacobian
(Eq.~\eqref{eq:log_jac}) is added to the log-posterior.

\paragraph{Prior bounds on reparametrized parameters.}
The sampled variables are constrained to the ranges
$\Omega_m\in(0.05,0.90)$, $\Omega_b\in(0.01,0.10)$,
$h\in(0.40,1.00)$, $m\in(0.50,2.50)$, $A>0$, and
$f_B\in[0,\,f_{B,\rm max}]$, where $f_{B,\rm max}$ corresponds to
the smallest admissible exponent $\alpha=1.5$ and keeps
$\sigma=m+3-\alpha$ safely below the $\sigma=4$ singularity.
After recovery the derived parameters must satisfy
$\gamma\in(0.01,5.0)$ and $\alpha\in(1.5,5.0)$; proposals failing
these bounds are rejected.  The complete prior specification is
summarized in Table~\ref{tab:priors}.

\begin{table}[h]
\centering
\caption{Flat prior ranges used by the reparametrized sampler.  The
absolute magnitude $M$ is sampled only for Pantheon$+$/SH0ES
combinations.  The prior on $f_B$ is model dependent because
$f_{B,\rm max}$ is evaluated at the current sampled $m$ and $h$.}
\label{tab:priors}
\begin{tabular}{lc}
\multicolumn{1}{c}{\bf Parameter} & \multicolumn{1}{c}{\bf Prior} \\
\hline
$\Omega_m$ & $(0.05,0.90)$\\
$\Omega_b$ & $(0.01,0.10)$ with $\Omega_b<\Omega_m$\\
$h$ & $(0.40,1.00)$\\
$M$ & $M<0$ when Pantheon$+$/SH0ES is active\\
$m$ & $(0.50,2.50)$\\
$A$ & $A>0$, with derived $\gamma\in(0.01,5.0)$\\
$f_B$ & $[0,f_{B,\rm max}(m,h)]$, with derived $\alpha\in(1.5,5.0)$\\
\end{tabular}
\end{table}

\subsection{MCMC setup and model comparison}
\label{sec:mcmc}

For each proposed parameter set, the implicit modified Friedmann
equation~\eqref{eq:Ea} is solved for $E^2(z)$ with a
Newton--Raphson scheme initialized at the $\Lambda$CDM solution,
following the numerical implementation of Paper I~\cite{paper-I};
proposals that reach a non-physical branch (vanishing or
negative $E^2$, or a negative effective dark-energy density) are
rejected.  The radiation density is not
sampled but derived from $T_{\text{CMB}}=2.726$~K and
$N_{\text{eff}}=3.046$ via
$\Omega_{r0} = (1+0.2271 N_{\text{eff}})\,2.469\times10^{-5}\,h^{-2}$.

Because the tightest bound reported below,
$|m-1|\lesssim10^{-4}$ (Sec.~\ref{sec:res_area}), might appear to rest
delicately on this solver, we quantify its accuracy.
Equation~\eqref{eq:Ea} is solved as a two-dimensional Newton iteration
in $(u,v)=(E^2,\Omega_{\rm de})$ with the analytic Jacobian and a
$\Lambda$CDM warm start, converging typically in three to four
iterations, and in at most five at any redshift entering the
likelihood for parameters within the reported posteriors.  Iteration
stops only when the relative step falls below $10^{-12}$ and both
residuals of Eq.~\eqref{eq:Ea}, measured relative to the scale of each
equation, fall below $10^{-10}$; the relative residual actually
achieved is $\lesssim5\times10^{-16}$ over the reported posteriors,
and remains below $10^{-12}$ throughout the prior volume.  As an
independent check, eliminating $v$ analytically reduces
Eq.~\eqref{eq:Ea} to a single scalar equation, which we solve both by
a bracketed Brent method and at $50$-digit precision; both agree with
the Newton result to better than one part in $10^{14}$.  For
$\beta=0$---including BTC-I, from which the $|m-1|\lesssim10^{-4}$
bound is obtained---this scalar equation inverts exactly,
\begin{equation}\label{eq:Eclosed}
  E(z) = \left[\frac{\Omega_{m0}(1+z)^3+\Omega_{r0}(1+z)^4
                    +\Omega_{\Lambda0}}{A}\right]^{1/(3-m)},
\end{equation}
and a single routine is used for all scenarios: for these it
reproduces Eq.~\eqref{eq:Eclosed} to $\lesssim10^{-14}$ over the full
integration range, so that bound does not rest on an iterative
solution at all.  Iteration is unavoidable only for the entanglement
scenarios ($\beta=1$), where $B\neq0$ gives a genuine two-term
equation.  This is far more accuracy than the bound on $m$ requires.
Since $A\simeq\gamma X^{1-m}$ with $\ln X\approx-140$
[Eqs.~\eqref{eq:X} and~\eqref{eq:amplification}], one has, at fixed
$\gamma$---the regime in which the $10^{-4}$ bound is obtained---
$\delta m=\delta\ln A/140$: numerical error propagates into the
exponent \emph{divided} by $140$, not multiplied by it.  The same
suppression applies to any multiplicative bias in the computed
effective dark-energy density, whatever its origin, so the achieved
residual corresponds to $\delta m\lesssim4\times10^{-18}$, more than
thirteen orders of magnitude below the quoted bound.  The
$(m,\gamma)\to(A,B)$ mapping is moreover closed-form and never passes
through the solver, which only returns $E^2(z)$ at fixed $A$ and
$B$.

The total log-likelihood is the sum of the contributions from the
active datasets.  The combined posterior is sampled using the
affine-invariant MCMC ensemble sampler
\texttt{emcee}~\cite{Foreman-Mackey:2012any} in the reparametrized
coordinates described in Sec.~\ref{sec:reparam}, with
$n_{\text{walkers}}=64$, a burn-in phase of 500 steps, and 5000
production steps.  Convergence is assessed via the integrated
autocorrelation time and the acceptance fraction; proposal moves
were tuned per dataset--model combination to keep the acceptance
fraction in the target range $0.2$--$0.5$.  Walkers are
initialized in a small ball around the $\Lambda$CDM point in the
sampled coordinates, with $f_B>0$ since the inversion
[Eq.~\eqref{eq:Newton_h}] requires $\ln f_B$ to be finite. Following completion
of the production run, the physical parameters $(\gamma,\,\alpha)$
are reconstructed for each individual sample and subsequently appended to the chain.
The supernova absolute magnitude $M$ is sampled jointly whenever the PP\&SH0ES likelihood is included in the analysis;
however, its inferred values are not reported in the tables.

Model comparison is performed through the Bayesian log-evidence
$\ln\mathcal{Z}$, computed from the MCMC chains using
\texttt{MCEvidence}~\cite{Heavens:2017afc}.  Evidence differences
$\Delta\ln\mathcal{Z} =
\ln\mathcal{Z}_{\rm model} - \ln\mathcal{Z}_{\rm BH\text{-}I}$
are quoted relative to the $\Lambda$CDM baseline (BH-I) and
interpreted on the Jeffreys scale:
$|\Delta\ln\mathcal{Z}|<1$ (inconclusive), $1$--$2.5$
(weak), $2.5$--$5$ (moderate), $>5$ (strong evidence) in favor
of the model with the higher evidence.

\section{Results and Discussion}
\label{sec:results}
\begin{table*}[t]
\scriptsize
\caption{Marginalized 68\% constraints for the $\Lambda$CDM baseline
(BH-I) and the area-law (Case~I) scenarios across all dataset
combinations.  Here $\Omega_m$ is the present matter density
parameter, $\Omega_b$ the baryon density parameter, and $h$ the
reduced Hubble constant.  For BH-II the sampled amplitude satisfies
$A=\gamma$ exactly at $m=1$; for BTC-II the derived MHR coupling
parameter $\gamma$ is reconstructed from $(A,m)$ via
Eq.~\eqref{eq:gamma_from_A}.
Uncertainties are quoted to two significant figures, rounded
where necessary so that no error bound is finer than its central
value.}
\label{tab:constraints_area}
\setlength{\tabcolsep}{3pt}
\resizebox{\textwidth}{!}{%
\begin{tabular}{lccccc}
\multicolumn{1}{c}{\bf Parameter} &
\multicolumn{1}{c}{\bf CMB+DESI} &
\multicolumn{1}{c}{\bf $+$CC} &
\multicolumn{1}{c}{\bf $+$PP\&SH0ES} &
\multicolumn{1}{c}{\bf $+$PP\&SH0ES$+$CC} &
\multicolumn{1}{c}{\bf DESI$+$PP\&SH0ES$+$CC} \\
\hline
\multicolumn{6}{l}{\textbf{BH-I ($\Lambda$CDM)}}\\
{\boldmath$\Omega_m       $} & $0.2969\pm 0.0039          $ & $0.2970\pm 0.0039          $ & $0.2931\pm 0.0037          $ & $0.2933\pm 0.0037          $ & $0.3034\pm 0.0077          $\\
{\boldmath$\Omega_b       $} & $0.04746\pm 0.00043        $ & $0.04748\pm 0.00043        $ & $0.04728\pm 0.00042        $ & $0.04730\pm 0.00042        $ & $0.0521\pm 0.0013          $\\
{\boldmath$h              $} & $0.6844\pm 0.0032          $ & $0.6843\pm 0.0032          $ & $0.6881\pm 0.0030          $ & $0.6879\pm 0.0031          $ & $0.7330\pm 0.0096          $\\
\hline
\multicolumn{6}{l}{\textbf{BH-II (free $A$)}}\\
{\boldmath$\Omega_m       $} & $0.2943\pm 0.0046          $ & $0.2945\pm 0.0045          $ & $0.2839\pm 0.0053          $ & $0.2841\pm 0.0053          $ & $0.251^{+0.053}_{-0.024}   $\\
{\boldmath$\Omega_b       $} & $0.04722\pm 0.00048        $ & $0.04724\pm 0.00048        $ & $0.04632\pm 0.00058        $ & $0.04634\pm 0.00058        $ & $0.0366^{+0.016}_{-0.0065} $\\
{\boldmath$h              $} & $0.6879^{+0.0036}_{-0.0050}$ & $0.6878^{+0.0036}_{-0.0049}$ & $0.7016\pm 0.0066          $ & $0.7014^{+0.0063}_{-0.0071}$ & $0.7331\pm 0.0094          $\\
{\boldmath$A              $} & $0.9917^{+0.0083}_{-0.0018}$ & $0.9918^{+0.0082}_{-0.0018}$ & $0.965\pm 0.015            $ & $0.966\pm 0.015            $ & $0.827^{+0.17}_{-0.062}    $\\
\hline
\multicolumn{6}{l}{\textbf{BTC-I (free $m$)}}\\
{\boldmath$\Omega_m       $} & $0.2946^{+0.0046}_{-0.0041}$ & $0.2946\pm 0.0044          $ & $0.2856\pm 0.0052          $ & $0.2859\pm 0.0052          $ & $0.286^{+0.020}_{-0.009}  $\\
{\boldmath$\Omega_b       $} & $0.04729\pm 0.00046        $ & $0.04729\pm 0.00046        $ & $0.04666\pm 0.00052        $ & $0.04668\pm 0.00052        $ & $0.0478^{+0.0048}_{-0.0015}$\\
{\boldmath$h              $} & $0.6872^{+0.0035}_{-0.0044}$ & $0.6872^{+0.0035}_{-0.0044}$ & $0.6979^{+0.0054}_{-0.0064}$ & $0.6975^{+0.0053}_{-0.0064}$ & $0.7324\pm 0.0096          $\\
{\boldmath$m              $} & $0.999945^{+0.000055}_{-0.000011}$ & $0.999944^{+0.000056}_{-0.000011}$ & $0.99979^{+0.00013}_{-0.00009}$ & $0.99979^{+0.00013}_{-0.00008}$ & $0.99957^{+0.00043}_{-0.00004}$\\
\hline
\multicolumn{6}{l}{\textbf{BTC-II (free $A$, $m$)}}\\
{\boldmath$\Omega_m       $} & $0.2933\pm 0.0045          $ & $0.2933\pm 0.0045          $ & $0.2867\pm 0.0047          $ & $0.2882^{+0.0049}_{-0.0041}$ & $0.268^{+0.032}_{-0.023}   $\\
{\boldmath$\Omega_b       $} & $0.04644^{+0.00098}_{-0.00063}$ & $0.04647^{+0.00098}_{-0.00064}$ & $0.04445^{+0.00078}_{-0.00089}$ & $0.04463\pm 0.00084        $ & $0.0450^{+0.0052}_{-0.0030}$\\
{\boldmath$h              $} & $0.6945^{+0.0052}_{-0.0089}$ & $0.6944^{+0.0050}_{-0.0087}$ & $0.7148\pm 0.0076          $ & $0.7130\pm 0.0076          $ & $0.7326\pm 0.0094          $\\
{\boldmath$A              $} & $0.9958^{+0.0091}_{-0.0044}$ & $0.9954^{+0.0097}_{-0.0045}$ & $0.995^{+0.016}_{-0.008}  $ & $0.996^{+0.015}_{-0.007}  $ & $0.884^{+0.12}_{-0.089}    $\\
{\boldmath$m              $} & $1.00095^{+0.00050}_{-0.0011}$ & $1.00092^{+0.00051}_{-0.0011}$ & $1.0030^{+0.0011}_{-0.0009}$ & $1.0029\pm 0.0010          $ & $0.9981\pm 0.0030          $\\
{\boldmath$\gamma$}~(derived) & $0.876^{+0.13}_{-0.066}    $ & $0.880^{+0.13}_{-0.066}    $ & $0.659^{+0.070}_{-0.11}    $ & $0.664^{+0.077}_{-0.099}   $ & $1.21^{+0.28}_{-0.47}      $\\
\end{tabular}}
\end{table*}

\begin{table*}[t]
\scriptsize
\caption{Marginalized 68\% constraints for the entanglement
(Case~II) scenarios across all dataset combinations.  One-sided
entries ($<$) are 68\% upper limits.  The derived entanglement
exponent $\alpha$ is recovered from $f_B$ via
Eq.~\eqref{eq:Newton_h}, and the derived $\gamma$ (BTC ENT-II) from
$(A,m)$ via Eq.~\eqref{eq:gamma_from_A}.
Precision conventions are as in
Table~\ref{tab:constraints_area}.}
\label{tab:constraints_ent}
\setlength{\tabcolsep}{3pt}
\resizebox{\textwidth}{!}{%
\begin{tabular}{lccccc}
\multicolumn{1}{c}{\bf Parameter} &
\multicolumn{1}{c}{\bf CMB+DESI} &
\multicolumn{1}{c}{\bf $+$CC} &
\multicolumn{1}{c}{\bf $+$PP\&SH0ES} &
\multicolumn{1}{c}{\bf $+$PP\&SH0ES$+$CC} &
\multicolumn{1}{c}{\bf DESI$+$PP\&SH0ES$+$CC} \\
\hline
\multicolumn{6}{l}{\textbf{BH ENT-I (free $f_B$)}}\\
{\boldmath$\Omega_m       $} & $0.2950\pm 0.0043          $ & $0.2951\pm 0.0043          $ & $0.2879\pm 0.0043          $ & $0.2886\pm 0.0043          $ & $0.3002\pm 0.0083          $\\
{\boldmath$\Omega_b       $} & $0.04713^{+0.00053}_{-0.00047}$ & $0.04714\pm 0.00051        $ & $0.04622\pm 0.00061        $ & $0.04641\pm 0.00057        $ & $0.0504^{+0.0022}_{-0.0015}$\\
{\boldmath$h              $} & $0.6882^{+0.0037}_{-0.0050}$ & $0.6881^{+0.0037}_{-0.0051}$ & $0.6996\pm 0.0059          $ & $0.6981^{+0.0052}_{-0.0062}$ & $0.7327\pm 0.0097          $\\
{\boldmath$f_B            $} & $< 0.00267                 $ & $< 0.00272                 $ & $0.0092^{+0.0041}_{-0.0051}$ & $0.0074^{+0.0033}_{-0.0045}$ & $< 0.00584                 $\\
{\boldmath$\alpha$}~(derived) & $2.0417^{+0.0054}_{-0.010} $ & $2.0415^{+0.0053}_{-0.010} $ & $2.0294^{+0.0021}_{-0.0050}$ & $2.0310^{+0.0023}_{-0.0052}$ & $2.0366^{+0.0056}_{-0.011} $\\
\hline
\multicolumn{6}{l}{\textbf{BH ENT-II (free $A$, $f_B$)}}\\
{\boldmath$\Omega_m       $} & $0.2938\pm 0.0045          $ & $0.2942\pm 0.0043          $ & $0.2857\pm 0.0050          $ & $0.2853\pm 0.0050          $ & $0.289^{+0.017}_{-0.009}  $\\
{\boldmath$\Omega_b       $} & $0.04689^{+0.00058}_{-0.00049}$ & $0.04693\pm 0.00051        $ & $0.04614\pm 0.00061        $ & $0.04607\pm 0.00062        $ & $0.0471^{+0.0051}_{-0.0017}$\\
{\boldmath$h              $} & $0.6906^{+0.0039}_{-0.0057}$ & $0.6902^{+0.0040}_{-0.0051}$ & $0.7020^{+0.0058}_{-0.0070}$ & $0.7029^{+0.0060}_{-0.0071}$ & $0.7330\pm 0.0099          $\\
{\boldmath$A              $} & $1.0031^{+0.0073}_{-0.012} $ & $1.0037^{+0.0080}_{-0.012} $ & $0.988^{+0.012}_{-0.015}   $ & $0.987\pm 0.015            $ & $0.977^{+0.035}_{-0.017}   $\\
{\boldmath$f_B            $} & $< 0.00548                 $ & $< 0.00574                 $ & $0.0070^{+0.0028}_{-0.0052}$ & $0.0076^{+0.0030}_{-0.0060}$ & $< 0.0115                  $\\
{\boldmath$\alpha$}~(derived) & $2.0364^{+0.0053}_{-0.0096}$ & $2.0362^{+0.0049}_{-0.0097}$ & $2.0319^{+0.0029}_{-0.0061}$ & $2.0314^{+0.0029}_{-0.0063}$ & $2.0313^{+0.0061}_{-0.0098}$\\
\hline
\multicolumn{6}{l}{\textbf{BTC ENT-I (free $m$, $f_B$)}}\\
{\boldmath$\Omega_m       $} & $0.2933\pm 0.0046          $ & $0.2935\pm 0.0046          $ & $0.2864^{+0.0053}_{-0.0047}$ & $0.2859\pm 0.0050          $ & $0.288^{+0.017}_{-0.011}   $\\
{\boldmath$\Omega_b       $} & $0.04665^{+0.00066}_{-0.00056}$ & $0.04667^{+0.00069}_{-0.00055}$ & $0.04594\pm 0.00060        $ & $0.04594\pm 0.00058        $ & $0.0469^{+0.0044}_{-0.0024}$\\
{\boldmath$h              $} & $0.6929^{+0.0048}_{-0.0063}$ & $0.6926^{+0.0047}_{-0.0063}$ & $0.7032^{+0.0056}_{-0.0066}$ & $0.7035^{+0.0055}_{-0.0064}$ & $0.733\pm 0.010            $\\
{\boldmath$m              $} & $1.000078^{+0.000080}_{-0.00015}$ & $1.000076^{+0.000076}_{-0.00014}$ & $1.00002^{+0.00015}_{-0.00019}$ & $1.00000^{+0.00015}_{-0.00018}$ & $0.99976^{+0.00035}_{-0.00018}$\\
{\boldmath$f_B            $} & $< 0.0103                  $ & $< 0.0101                  $ & $0.0132^{+0.0049}_{-0.0098}$ & $0.0124^{+0.0051}_{-0.0090}$ & $< 0.0115                  $\\
{\boldmath$\alpha$}~(derived) & $2.0322^{+0.0046}_{-0.0099}$ & $2.0323^{+0.0048}_{-0.0097}$ & $2.0272^{+0.0031}_{-0.0058}$ & $2.0275^{+0.0027}_{-0.0058}$ & $2.0311^{+0.0045}_{-0.0090}$\\
\hline
\multicolumn{6}{l}{\textbf{BTC ENT-II (free $A$, $m$, $f_B$)}}\\
{\boldmath$\Omega_m       $} & $0.2927\pm 0.0049          $ & $0.2929\pm 0.0046          $ & $0.2876^{+0.0049}_{-0.0044}$ & $0.2877\pm 0.0046          $ & $0.276^{+0.029}_{-0.023}   $\\
{\boldmath$\Omega_b       $} & $0.0463^{+0.0011}_{-0.0008}$ & $0.0463^{+0.0011}_{-0.0007}$ & $0.0447\pm 0.0010          $ & $0.04478^{+0.00097}_{-0.0011}$ & $0.0468^{+0.0040}_{-0.0024}$\\
{\boldmath$h              $} & $0.6963^{+0.0059}_{-0.010} $ & $0.6956^{+0.0060}_{-0.0099}$ & $0.7121\pm 0.0092          $ & $0.7119\pm 0.0090          $ & $0.7314^{+0.0097}_{-0.011} $\\
{\boldmath$A              $} & $1.015^{+0.011}_{-0.022}   $ & $1.017^{+0.011}_{-0.023}   $ & $1.010\pm 0.017            $ & $1.010\pm 0.017            $ & $0.925^{+0.089}_{-0.074}   $\\
{\boldmath$m              $} & $1.00045^{+0.00080}_{-0.0012}$ & $1.00038^{+0.00089}_{-0.0012}$ & $1.0021\pm 0.0012          $ & $1.0021\pm 0.0012          $ & $0.9977^{+0.0028}_{-0.0023}$\\
{\boldmath$f_B            $} & $< 0.0114                  $ & $< 0.0120                  $ & $< 0.00977                 $ & $0.0078^{+0.0020}_{-0.0074}$ & $< 0.0128                  $\\
{\boldmath$\gamma$}~(derived) & $0.96\pm 0.14              $ & $0.97\pm 0.15              $ & $0.76^{+0.11}_{-0.13}      $ & $0.76^{+0.11}_{-0.14}      $ & $1.34^{+0.24}_{-0.48}      $\\
{\boldmath$\alpha$}~(derived) & $2.0308^{+0.0045}_{-0.0087}$ & $2.0306^{+0.0050}_{-0.0089}$ & $2.0316^{+0.0033}_{-0.0074}$ & $2.0316^{+0.0037}_{-0.0072}$ & $2.0301^{+0.0041}_{-0.0083}$\\
\end{tabular}}
\end{table*}


\begin{table*}[t]
\scriptsize
\caption{Bayesian log-evidence differences
$\Delta\ln\mathcal{Z}=\ln\mathcal{Z}_{\rm model}
-\ln\mathcal{Z}_{\Lambda{\rm CDM}}$, computed from the MCMC
chains with \texttt{MCEvidence}~\cite{Heavens:2017afc}, for all
extended scenarios and dataset combinations.  Negative values
favor $\Lambda$CDM.  The first row lists the $\Lambda$CDM (BH-I)
reference $\ln\mathcal{Z}_{\Lambda{\rm CDM}}$; the
\texttt{MCEvidence} statistical uncertainties on the individual
$\ln\mathcal{Z}$ values are $\lesssim0.45$, so the uncertainty on
$\Delta\ln\mathcal{Z}$ is $\lesssim0.5$.
Values are quoted to one decimal place, consistent with this
uncertainty.  The last column summarizes the strength of the evidence
\emph{against} the extension on the Jeffreys scale defined in
Sec.~\ref{sec:mcmc}, and gives the range spanned by the five dataset
combinations.}
\label{tab:evidence}
\resizebox{\textwidth}{!}{%
\begin{tabular}{lccccc|c}
\multicolumn{1}{c}{\bf Model} &
\multicolumn{1}{c}{\bf CMB+DESI} &
\multicolumn{1}{c}{\bf $+$CC} &
\multicolumn{1}{c}{\bf $+$PP\&SH0ES} &
\multicolumn{1}{c}{\bf $+$PP\&SH0ES$+$CC} &
\multicolumn{1}{c}{\bf DESI$+$PP\&SH0ES$+$CC} &
\multicolumn{1}{c}{\bf Evidence against extension} \\
\hline
$\ln\mathcal{Z}_{\Lambda{\rm CDM}}$ & $-25.32\pm0.12$ & $-32.50\pm0.06$ & $-807.63\pm0.15$ & $-814.83\pm0.08$ & $-795.25\pm0.08$ & \textbf{---} \\
\hline
BH-II      & $-4.9$  & $-4.8$  & $-1.0$ & $-1.0$ & $-1.8$  & \textbf{weak--moderate}     \\
BTC-I      & $-9.9$  & $-9.8$  & $-6.9$ & $-6.9$ & $-8.1$  & \textbf{strong}             \\
BTC-II     & $-11.3$ & $-11.3$ & $-4.4$ & $-4.1$ & $-8.8$  & \textbf{moderate--strong}   \\
BH ENT-I   & $-6.3$  & $-6.2$  & $-2.7$ & $-3.2$ & $-5.7$  & \textbf{moderate--strong}   \\
BH ENT-II  & $-10.6$ & $-10.6$ & $-6.8$ & $-6.8$ & $-9.2$  & \textbf{strong}             \\
BTC ENT-I  & $-14.7$ & $-14.8$ & $-9.1$ & $-9.4$ & $-13.2$ & \textbf{strong}             \\
BTC ENT-II & $-16.1$ & $-16.3$ & $-8.7$ & $-8.7$ & $-14.3$ & \textbf{strong}             \\
\end{tabular}}
\end{table*}

\begin{figure*}[t]
\centering

\begin{subfigure}{0.48\textwidth}
    \centering
    \includegraphics[width=\linewidth]{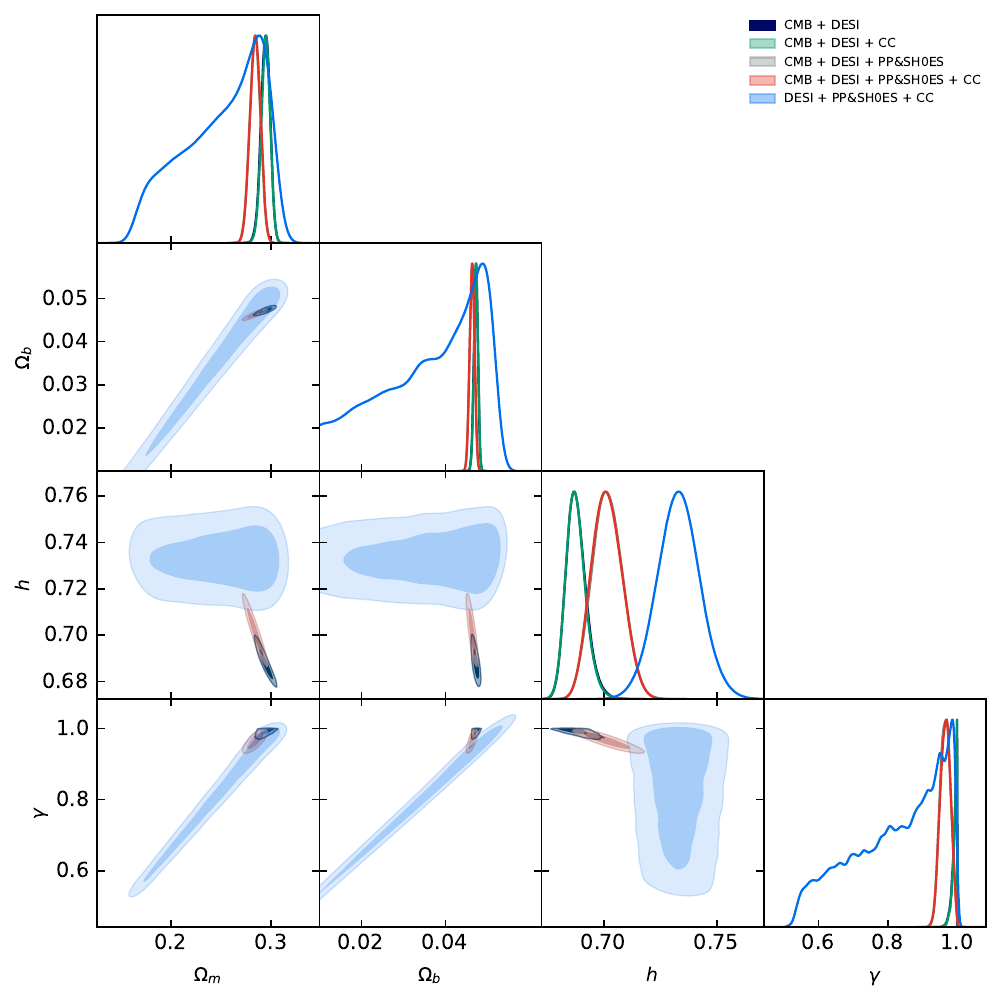}
    \caption{BH-II ($A$ free)}
\end{subfigure}
\hfill
\begin{subfigure}{0.48\textwidth}
    \centering
    \includegraphics[width=\linewidth]{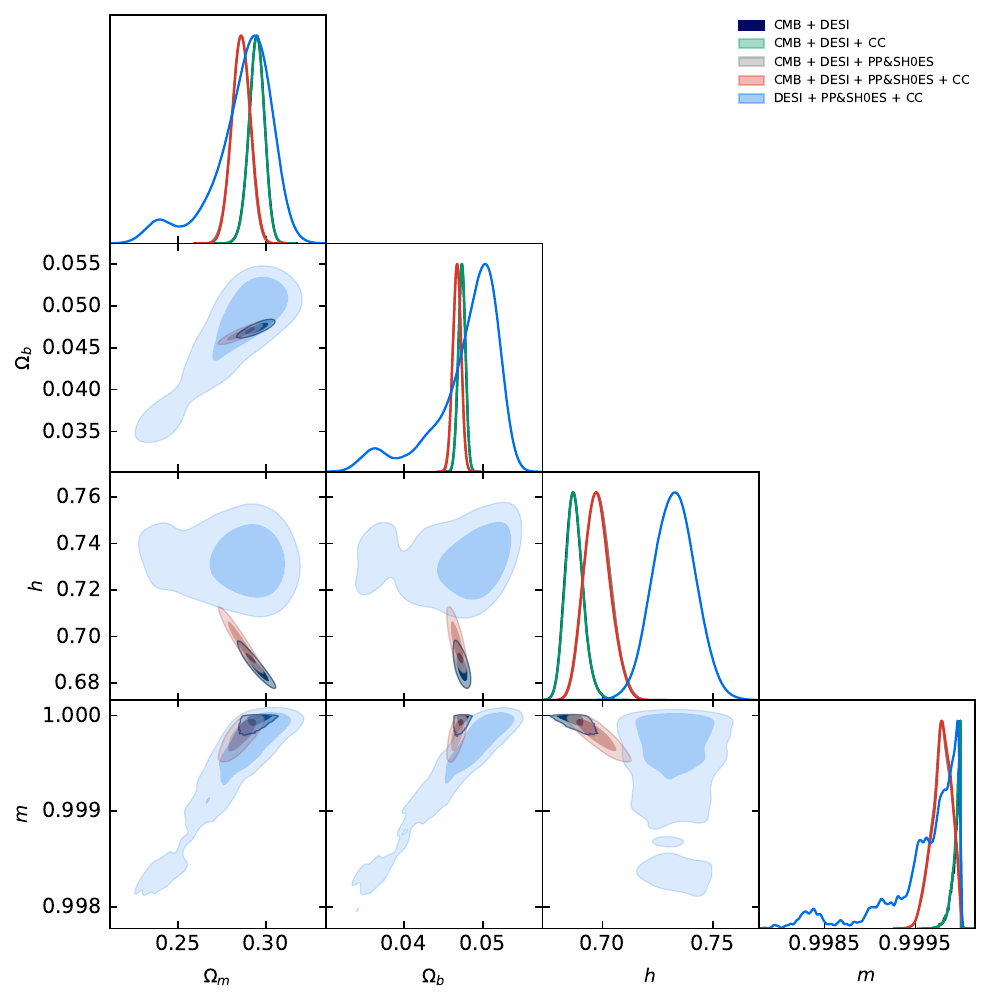}
    \caption{BTC-I ($m$ free)}
\end{subfigure}

\caption{
Marginalized 68\% and 95\% confidence contours for the
single-parameter extensions BH-II and BTC-I.
Five dataset combinations are listed in the legend, but fewer
distinct curves are visible in most panels: the CMB+DESI and
CMB+DESI+CC contours overlap almost exactly, as do the
CMB+DESI+PP\&SH0ES and CMB+DESI+PP\&SH0ES+CC contours.  This is the
expected consequence of the fact that adding cosmic chronometers does
not shift the central values (Sec.~\ref{sec:res_baseline}); the
corresponding pairs of columns in
Table~\ref{tab:constraints_area} agree to within a small fraction of
$1\sigma$.
}
\label{fig:tri_1}
\end{figure*}

\begin{figure*}[t]
\centering

\begin{subfigure}{0.48\textwidth}
    \centering
    \includegraphics[width=\linewidth]{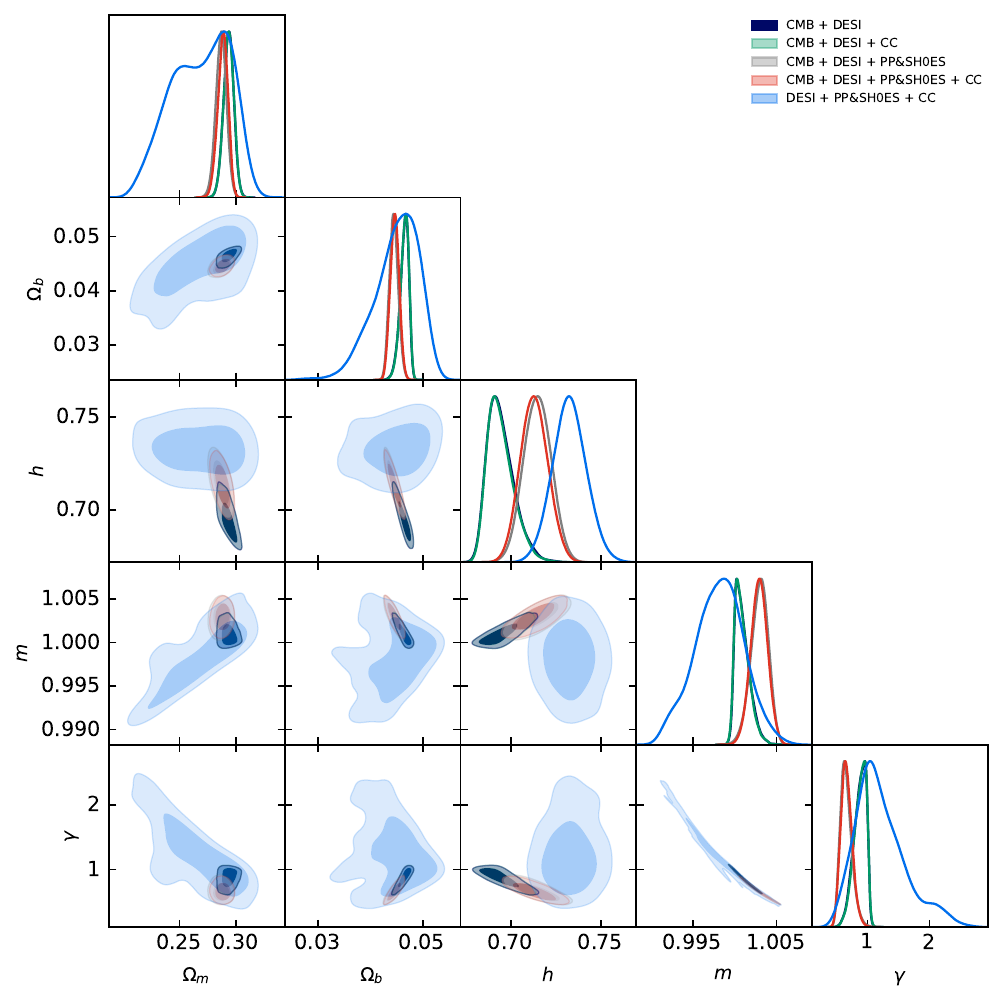}
    \caption{BTC-II ($A$, $m$ free)}
\end{subfigure}
\hfill
\begin{subfigure}{0.48\textwidth}
    \centering
    \includegraphics[width=\linewidth]{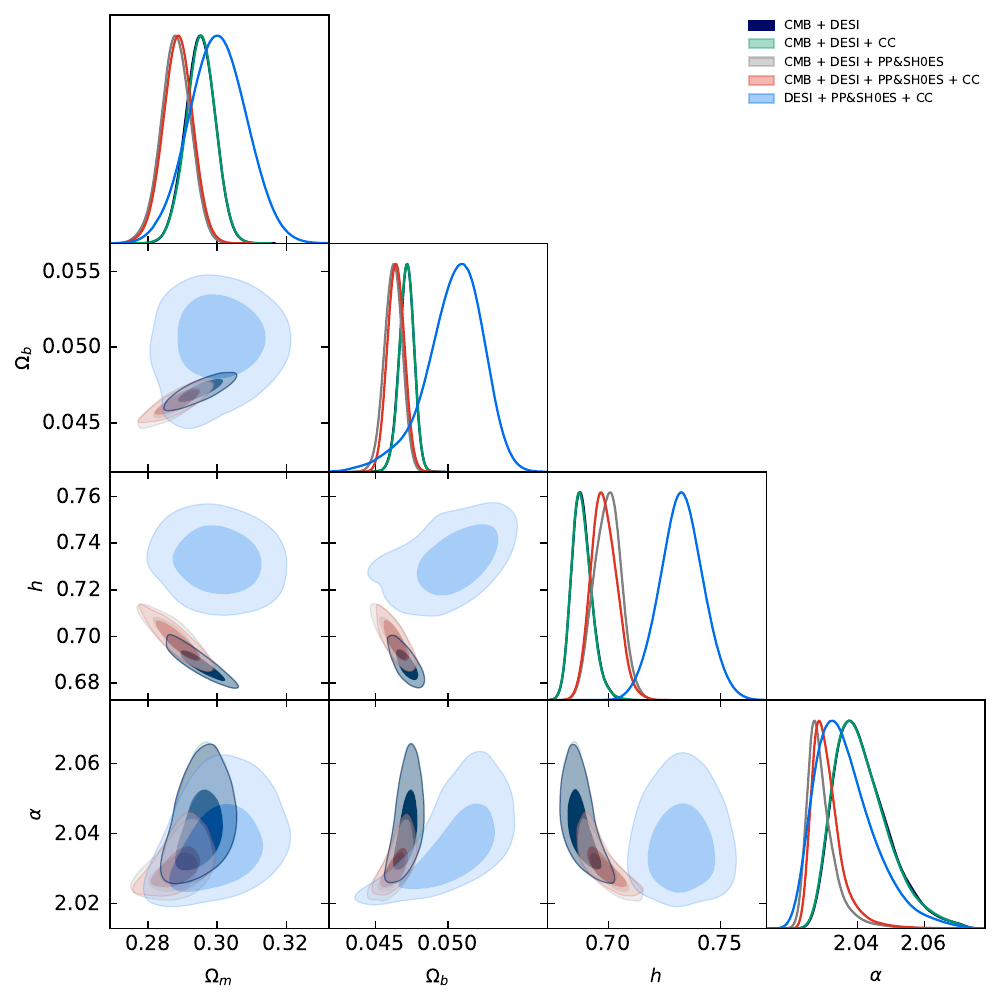}
    \caption{BH ENT-I ($f_B$ free)}
\end{subfigure}

\caption{
Marginalized posterior contours for the BTC-II and BH ENT-I
scenarios.
As in Fig.~\ref{fig:tri_1}, the CMB+DESI and CMB+DESI+CC
contours overlap almost exactly, as do the CMB+DESI+PP\&SH0ES and
CMB+DESI+PP\&SH0ES+CC contours, so fewer than five distinct curves are
visible in most panels.
}
\label{fig:tri_2}
\end{figure*}

\begin{figure*}[t]
\centering

\begin{subfigure}{0.48\textwidth}
    \centering
    \includegraphics[width=\linewidth]{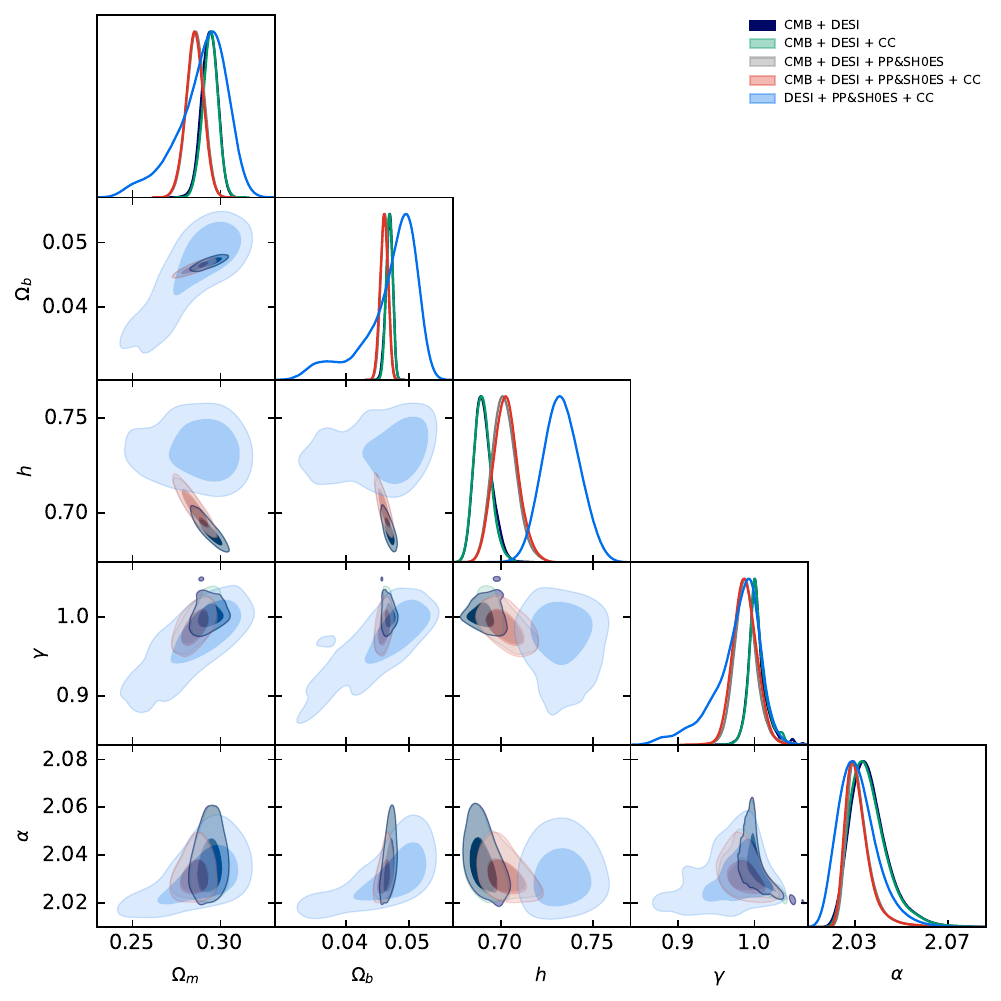}
    \caption{BH ENT-II ($A$, $f_B$ free)}
\end{subfigure}
\hfill
\begin{subfigure}{0.48\textwidth}
    \centering
    \includegraphics[width=\linewidth]{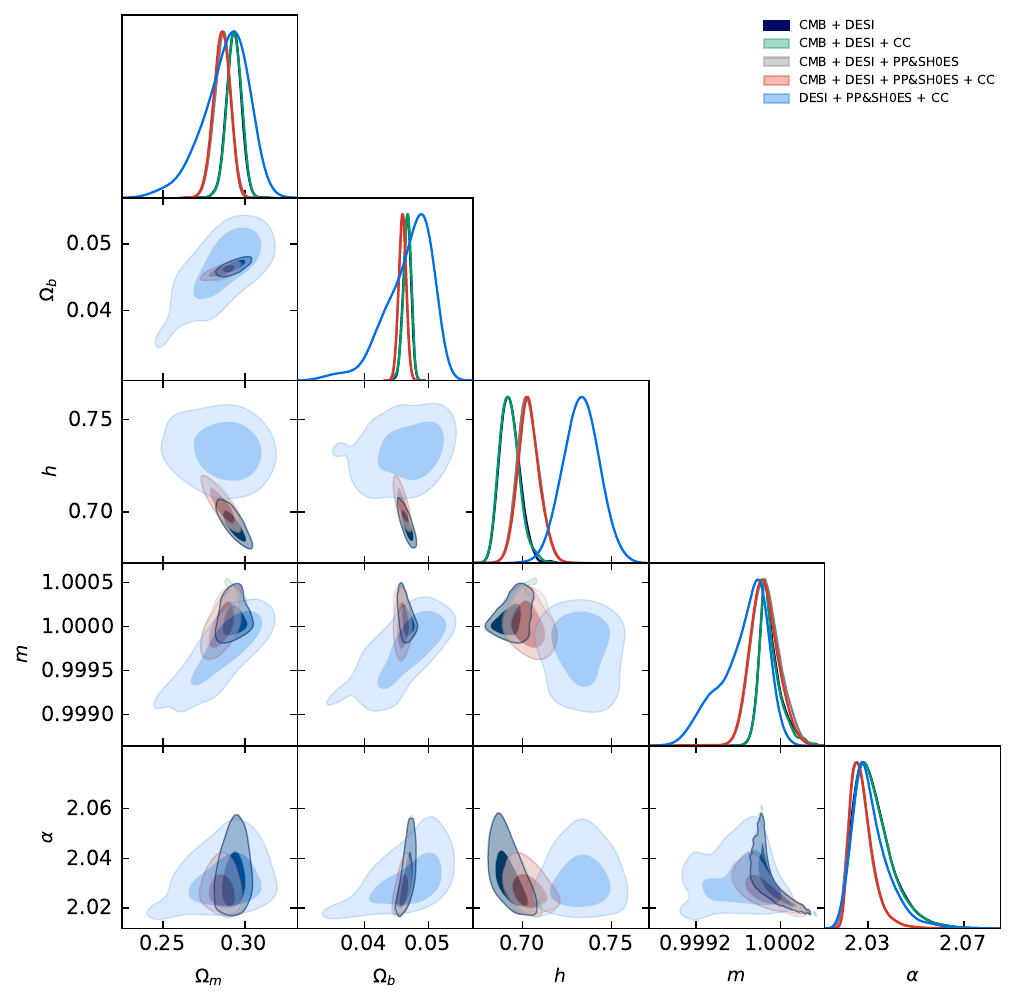}
    \caption{BTC ENT-I ($m$, $f_B$ free)}
\end{subfigure}

\caption{
Marginalized posterior contours for the BH ENT-II and
BTC ENT-I scenarios.
As in Fig.~\ref{fig:tri_1}, the CMB+DESI and CMB+DESI+CC
contours overlap almost exactly, as do the CMB+DESI+PP\&SH0ES and
CMB+DESI+PP\&SH0ES+CC contours, so fewer than five distinct curves are
visible in most panels.
}
\label{fig:tri_3}
\end{figure*}

\begin{figure*}[t]
\centering

\begin{subfigure}{0.48\textwidth}
    \centering
    \includegraphics[width=\linewidth]{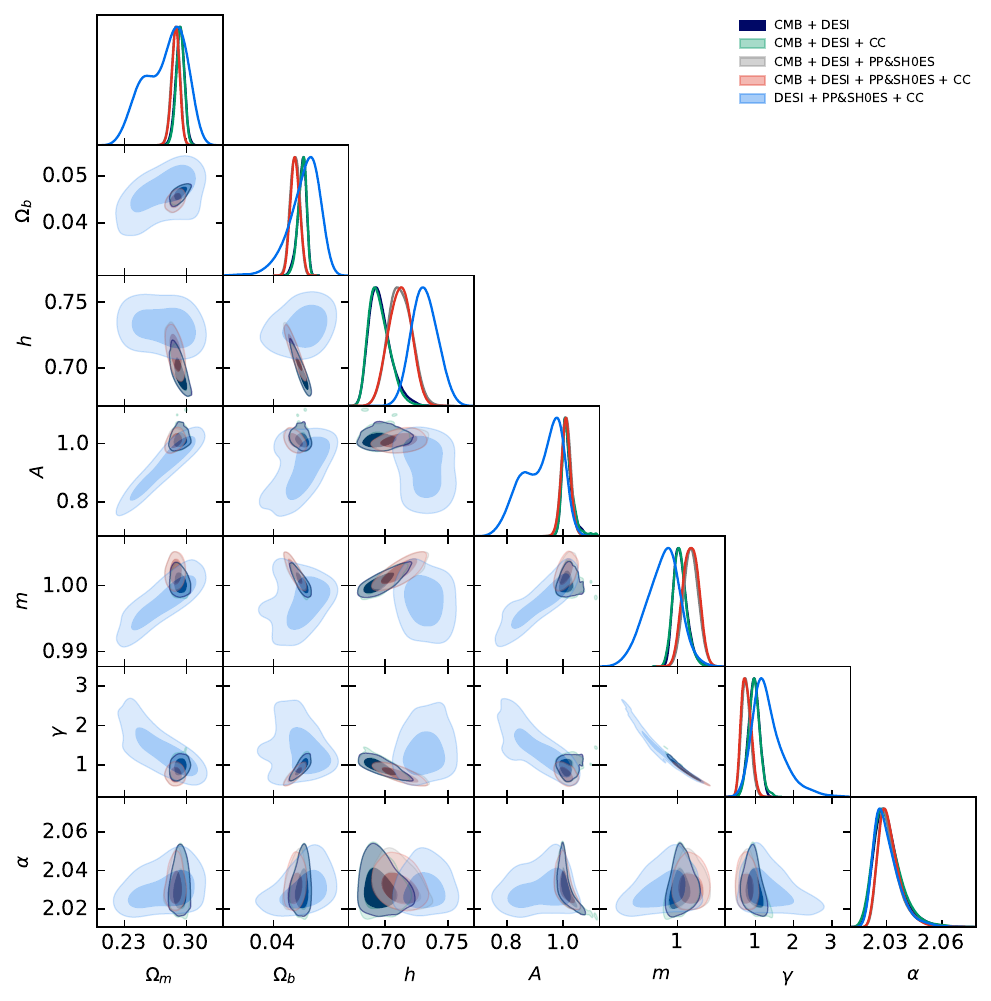}
    \caption{BTC ENT-II ($A$, $m$, $f_B$ free)}
\end{subfigure}
\hfill
\begin{subfigure}{0.48\textwidth}
    \centering
    \includegraphics[width=\linewidth]{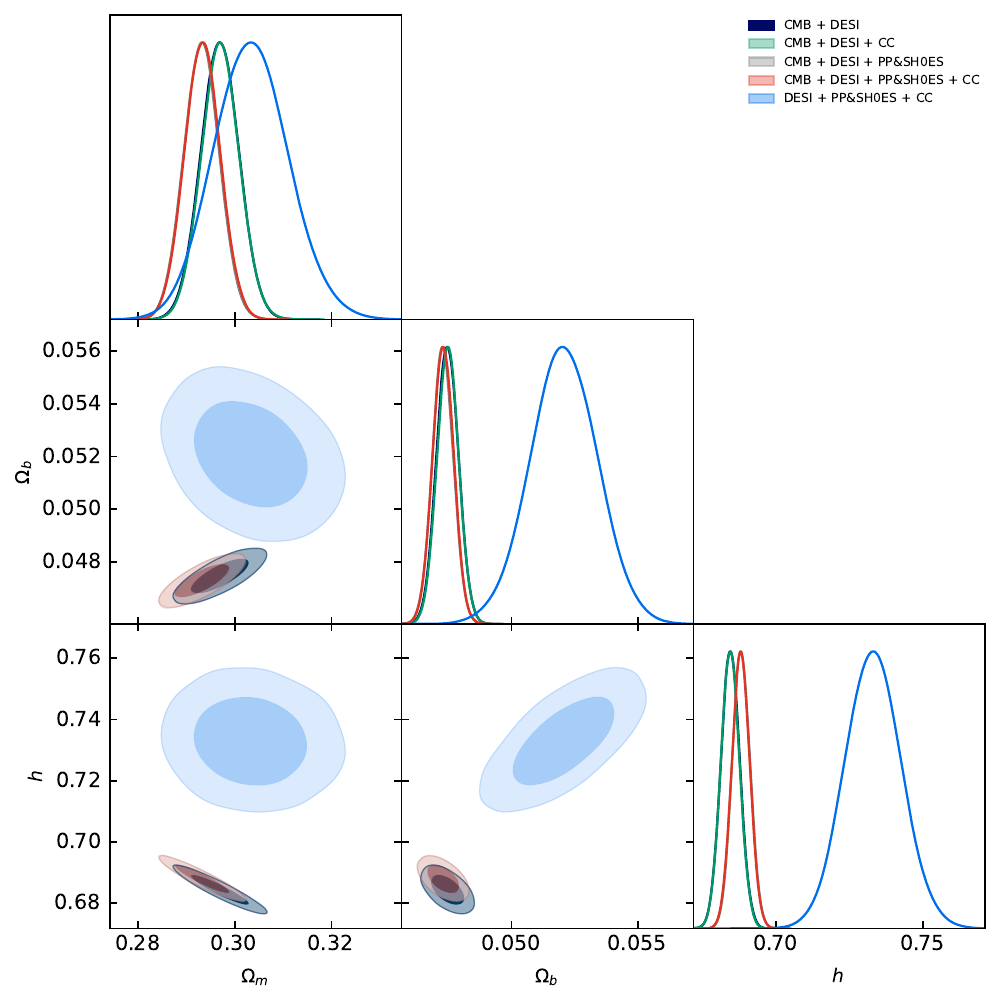}
    \caption{BH-I ($\Lambda$CDM reference)}
\end{subfigure}

\caption{
Marginalized posterior contours for the most general
extended MHR scenario, BTC ENT-II, and the BH-I
($\Lambda$CDM) reference cosmology.
As in Fig.~\ref{fig:tri_1}, the CMB+DESI and CMB+DESI+CC
contours overlap almost exactly, as do the CMB+DESI+PP\&SH0ES and
CMB+DESI+PP\&SH0ES+CC contours, so fewer than five distinct curves are
visible in most panels.
}
\label{fig:tri_4}
\end{figure*}

The primary results are presented for the scenarios summarized in
Table~\ref{tab:cases}, with the marginalized 68\%
confidence-level constraints reported in
Tables~\ref{tab:constraints_area} and~\ref{tab:constraints_ent}
for five data combinations: CMB+DESI, CMB+DESI+CC,
CMB+DESI+PP\&SH0ES, the full combination CMB+DESI+PP\&SH0ES+CC,
and a CMB-free combination DESI+PP\&SH0ES+CC.  The CMB-free
combination retains the SH0ES Cepheid calibration and is
therefore anchored by the local distance ladder; its comparison
with the CMB-inclusive chains isolates the tension between the
early-universe and local calibrations of $H_0$, independently of
the CMB compressed likelihood.  Full triangle plots for all
scenarios are collected in
Figs.~\ref{fig:tri_1}--\ref{fig:tri_4}.

Three principal conclusions can be drawn from this analysis.
(\textit{i})~The entropy exponent is tightly constrained around its Bekenstein--Hawking value, with $|m-1|\lesssim\mathcal{O}(10^{-4})$ at $\gamma=1$.  Under the algebraic mapping $\Delta=m-1$, BTC-I gives $\Delta=(-5.5^{+5.5}_{-1.1})\times10^{-5}$ (CMB+DESI) and $\Delta=(-2.1^{+1.3}_{-0.8})\times10^{-4}$ (Full Data), both on the negative side.  The BTC-II and BTC ENT-II Full-Data constraints are positive, whereas BTC-II with CMB+DESI, both BTC ENT-I constraints, and BTC ENT-II with CMB+DESI cross $\Delta=0$.  No prior $\Delta\in[0,1]$ (equivalently $m\in[1,2]$) was imposed in these chains, so the negative and zero-crossing intervals must be retained rather than truncated.
(\textit{ii})~Allowing the MHR coupling or the entanglement amplitude to vary leads to an inferred Hubble constant of $h\simeq0.70$–$0.71$, thereby reducing the CMB–SH0ES discrepancy from ${\sim}4\sigma$ to ${\sim}1.2$–$2.6\sigma$, although none of the considered scenarios fully resolves the tension. (\textit{iii})~The Bayesian evidence disfavors
every extension ($-16\lesssim\Delta\ln\mathcal{Z}\lesssim-1$):
the apparent preference for modified entropy when SH0ES is
included is an absorption of the Hubble tension, not evidence
for modified horizon thermodynamics.  The remainder of this
section details these results: the $\Lambda$CDM baseline
(Sec.~\ref{sec:res_baseline}), the area-law and entanglement
families (Secs.~\ref{sec:res_area} and~\ref{sec:res_ent}), the
Hubble tension across scenarios (Sec.~\ref{subsec:H0_summary}),
Bayesian model comparison (Sec.~\ref{subsec:comparison},
Table~\ref{tab:evidence}), and a comparison with previous
studies (Sec.~\ref{sec:lit_comparison}).

\subsection{Baseline: BH-I ($\Lambda$CDM) and dataset consistency}
\label{sec:res_baseline}

The BH-I scenario corresponds exactly to $\Lambda$CDM ($m=1$,
$\gamma=1$, $\beta=0$) and sets the benchmark the extended models
must improve upon: $h = 0.6881\pm0.0030$ from
CMB+DESI+PP\&SH0ES---a ${\sim}3.9\sigma$ tension with the SH0ES
value $H_0 = 73.04 \pm 1.04$~km\,s$^{-1}$\,Mpc$^{-1}$---versus
$h = 0.7330 \pm 0.0096$ in the CMB-free chain
(Table~\ref{tab:constraints_area}, first block).  The matter and
baryon densities agree with the Planck~2018 best
fit~\cite{Planck:2018vyg} in all CMB-inclusive combinations, and
adding cosmic chronometers (CC) does not shift the central
values, confirming the mutual consistency of the geometric
probes.  The CMB-free chain locks onto the SH0ES value because,
without the Planck distance priors that anchor the sound horizon,
the constraint on $H_0$ comes entirely from the Cepheid
calibration and the shape of the BAO and supernova distance
ratios.  This is the CMB-vs-local-ladder discrepancy that the
extended entropy models attempt to absorb.

\subsection{Area-law extensions (Case I): the coupling $\gamma$ and the exponent $m$}
\label{sec:res_area}

The area-law (Case I) extensions BH-II, BTC-I, and BTC-II
(Table~\ref{tab:constraints_area}; posterior distributions in
Figs.~\ref{fig:tri_1} and~\ref{fig:tri_2}) yield two of the main
results: the MHR coupling $\gamma$ absorbs part of the Hubble
tension, raising $h$ to $0.7016$--$0.7148$ at the cost of a
coupling below unity, while the entropy exponent is pinned to
$|m-1|\lesssim10^{-4}$ by the scale-hierarchy amplification
derived below.

Freeing the MHR coupling parameter $\gamma$ while keeping $m=1$
and $\beta=0$ yields the BH-II model, whose Friedmann equation
reduces to $\gamma H^2 = (8\pi
G/3c^2)(\varepsilon_m+\varepsilon_r) + \Lambda c^2/3$---at the
homogeneous level an effective $G_{\rm eff}=G/\gamma$ and
$\Lambda_{\rm eff}=\Lambda/\gamma$.  The sampled parameter $A$
(equal to $\gamma$ exactly at $m=1$, Sec.~\ref{sec:reparam}) is
consistently displaced below unity:
$A = 0.9917^{+0.0083}_{-0.0018}$ from CMB+DESI alone and
$A = 0.965 \pm 0.015$ when PP\&SH0ES are included, with $h$
rising simultaneously to $0.7016 \pm 0.0066$: a reduced $\gamma$
shifts $\Lambda_{\rm eff}$ upward and accommodates the higher
$H_0$ preferred by the Cepheid distances.  This partially
alleviates the Hubble tension but does not resolve it, leaving a
residual ${\sim}2.3\sigma$ discrepancy with SH0ES.  In the
CMB-free chain $A$ broadens to $0.827^{+0.17}_{-0.062}$ while
$h = 0.7331 \pm 0.0094$ matches the SH0ES anchor: without the
CMB acoustic scale, the data cannot separate a rescaled $\gamma$
from a shifted $(\Omega_{\Lambda 0},H_0)$ combination.  The
SH0ES-driven preference $A=\gamma\simeq0.965$ lies below the
background-viability window $0.981\le\gamma\le1$ of
Paper I~\cite{paper-I}, derived there at fixed fiducial
parameters; the joint fit accommodates it by adjusting $\Omega_m$
and $h$ self-consistently.

The BTC-I scenario keeps $\gamma=1$ and $\beta=0$ and frees only
the entropy exponent, with remarkably tight results:
$m = 0.999945^{+0.000055}_{-0.000011}$ from CMB+DESI and
$m = 0.99979^{+0.00013}_{-0.00008}$ from the full combination,
always consistent with $m=1$ well below the $1\sigma$ level.
This precision originates in the hierarchy
$X=\ell_{\rm Pl} H_0/c \sim 10^{-61}$ [Eq.~\eqref{eq:X}]: near
$m=1$, $A \simeq \gamma\,X^{1-m}$ [Eq.~\eqref{eq:AB}], so a
fractional departure $\delta m$ from unity shifts $A$ by
\begin{equation}\label{eq:amplification}
\frac{\delta A}{A} \approx |\delta m|\,\ln\!\left(\frac{c}
  {\ell_{\rm Pl} H_0}\right) \approx 140\,|\delta m|,
\end{equation}
an enhancement we refer to as the \emph{scale-hierarchy
amplification}.  Since
$\Omega_{\Lambda 0} = A - (\Omega_{m0}+\Omega_{r0})$
[Eq.~\eqref{eq:OmLambda}] is constrained at the sub-percent level
by CMB+BAO, Eq.~\eqref{eq:amplification} immediately implies
$|\delta m|\lesssim 10^{-4}$: the tight bound on $m$ reflects the
amplification of the Planck-to-Hubble hierarchy onto
$\Omega_{\Lambda 0}$, not a direct observational sensitivity to
the fractal geometry of the cosmological horizon.

In the Barrow entropy framework~\cite{Barrow:2020tzx}, the
fractal structure of the horizon is parametrized by
$\Delta=m-1$, physically restricted to $\Delta\in[0,1]$
(Sec.~\ref{sec:lit_comparison}).  Our prior-unrestricted BTC-I
posteriors give $\Delta=(-5.5^{+5.5}_{-1.1})\times10^{-5}$
(CMB+DESI) and $\Delta=(-2.1^{+1.3}_{-0.8})\times10^{-4}$
(Full Data), so both lie at or below the physical Barrow
boundary $\Delta=0$ at 68\% C.L.\ (the Full-Data interval strictly
below, the CMB+DESI interval reaching $\Delta=0$ exactly at its
upper edge).  These values are exact translations of the sampled MHR
exponent, not Barrow-prior constraints, because no restriction
$m\in[1,2]$ was imposed.  In the Tsallis--Cirto
framework~\cite{Tsallis:2012js}, $\delta=(m+1)/2$ is not
domain-restricted, and the same translation applies directly.
In the CMB-free chain the constraint relaxes to
$m = 0.99957^{+0.00043}_{-0.00004}$, still consistent with $m=1$
at $1\sigma$, confirming that the tight bound on the entropy
exponent is driven by the late-universe geometric data and is not
an artifact of the CMB prior alone.

When both $m$ and $\gamma$ are freed simultaneously (BTC-II),
the individual posteriors broaden substantially, revealing a
pronounced degeneracy: near $m=1$ only the combination
$A=(2m\gamma/(3-m))\,X^{1-m}$ controls
$\Omega_{\Lambda0}=A-(\Omega_{m0}+\Omega_{r0})$, so a joint
change $(\delta m,\delta\gamma)$ satisfying the degeneracy
condition~\eqref{eq:degeneracy_condition} leaves the expansion
history indistinguishable from $\Lambda$CDM along a steep ridge
in the $m$--$\gamma$ plane (slope
$\delta\gamma/\delta m \approx -140$).  We therefore sample
$(A,m)$ with the Jacobian correction of Eq.~\eqref{eq:log_jac},
as described in Sec.~\ref{sec:reparam}.

With CMB+DESI, the posterior peak lies at
$m = 1.00095^{+0.00050}_{-0.0011}$ and
$A = 0.9958^{+0.0091}_{-0.0044}$, consistent with
both $m=1$ and $A=1$ within $1\sigma$.
Including PP\&SH0ES yields
$m = 1.0030^{+0.0011}_{-0.0009}$ and
$A = 0.995^{+0.016}_{-0.008}$.
Translated into the Barrow language,
$\Delta = m-1 = 0.0030^{+0.0011}_{-0.0009}$ with PP\&SH0ES, a
nominally $\sim 3\sigma$ non-zero value.  However, this
apparent signal should be attributed to the $m$--$\gamma$
degeneracy: the sampled combination $A$ is well-constrained
near unity, while $m$ and $\gamma$ individually are degenerate
along the steep ridge described above.
The Hubble constant rises to $h = 0.7148\pm 0.0076$ with PP\&SH0ES,
the largest shift seen in the CMB-inclusive analysis.  In the
CMB-free chain, $A = 0.884^{+0.12}_{-0.089}$ and
$m = 0.9981\pm 0.0030$, both consistent with $\Lambda$CDM.

The derived coupling (Table~\ref{tab:constraints_area}, last row
of the BTC-II block) makes the ridge explicit: with PP\&SH0ES,
$\gamma = 0.659^{+0.070}_{-0.11}$ at $m\simeq1.003$, tracing the
continuation of the viability ridge identified in
Paper I~\cite{paper-I} ($0.864\le\gamma\le0.871$ at $m=1.001$);
the two analyses probe the same degenerate direction.

\subsection{Entanglement corrections (Case II): the amplitude $f_B$}
\label{sec:res_ent}

The ENT scenarios activate the subleading entropy term
($\beta=1$), constrained through the well-conditioned amplitude
$f_B\equiv B/[(3-m)A]$ [Eq.~\eqref{eq:fB}], from which the
physical exponent $\alpha$ is recovered via
Eq.~\eqref{eq:Newton_h} (Table~\ref{tab:constraints_ent};
posterior distributions in
Figs.~\ref{fig:tri_2}--\ref{fig:tri_4}).  The key result of
Case II is a ${\sim}2\sigma$ preference for a nonzero
entanglement amplitude, $f_B \approx 0.009$, that appears only
when the SH0ES calibration is included.

In the simplest case, BH ENT-I ($m=\gamma=1$), the posterior on
$f_B$ is consistent with zero in the CMB-only combinations
($f_B < 0.00267$ at 68\% C.L.\ from CMB+DESI), whereas including
PP\&SH0ES yields $f_B = 0.0092^{+0.0041}_{-0.0051}$ (PP\&SH0ES)
and $f_B = 0.0074^{+0.0033}_{-0.0045}$ (full combination).  The
corresponding Hubble-constant shift is $\Delta h \simeq 0.011$
(from $0.6882$ to $0.6996\pm 0.0059$), compared with only
$\approx 0.003$ for $\Lambda$CDM on the same data.  Physically,
a positive $f_B$ switches on the subleading term
$\propto(r_a/\ell_{\rm Pl})^{\sigma}$ in the
entropy~\eqref{eq:entropy_def}, which raises the effective
dark-energy contribution at high redshift and lets a higher
$H_0$ fit the same angular distances.  In the CMB-free chain,
$f_B < 0.00584$, consistent with zero.

The remaining Case II scenarios follow the same pattern, with
their constraints fully reported in
Table~\ref{tab:constraints_ent}.  In BH ENT-II, freeing $A$
alongside $f_B$ dilutes the entanglement signal to a weak
preference ($f_B = 0.0070^{+0.0028}_{-0.0052}$ with PP\&SH0ES):
the $A$--$f_B$ degeneracy lets the overall coupling absorb the
dominant effect on the dark-energy density; tests with wider
priors confirm the bounds are not a prior artifact.  In
BTC ENT-I the exponent remains as tightly constrained as in
BTC-I---the same scale-hierarchy amplification---while $f_B$
mirrors BH ENT-I with a larger amplitude
($0.0132^{+0.0049}_{-0.0098}$) and
$h = 0.7032^{+0.0056}_{-0.0066}$ (PP\&SH0ES).  The most general
scenario, BTC ENT-II, reaches $h = 0.7121\pm 0.0092$, second
only to BTC-II, with $m$ mildly above unity as a projection of
the $m$--$\gamma$ degeneracy; in the CMB-free chain all
extension parameters are consistent with $\Lambda$CDM.

Across the ENT scenarios, the robust statements are: the MHR
exponent remains within $\mathcal{O}(10^{-3})$ of $m=1$ when
$\gamma$ is free (dominated by the $m$--$\gamma$ degeneracy); the derived
coupling deviates from unity only along the degeneracy
ridge---at 95\% C.L., $\gamma=0.66^{+0.19}_{-0.17}$ (BTC-II) and
$\gamma=0.76^{+0.24}_{-0.22}$ (BTC ENT-II)---and the derived
exponent clusters at $\alpha\simeq2.02$--$2.04$ whenever the
amplitude is nonzero (Table~\ref{tab:constraints_ent}, derived
rows).  Because $\alpha$ is not identifiable in the exact
$f_B=0$ limit, the amplitude $f_B$ is the physically more robust
quantity.

\subsection{Summary of Hubble tension across scenarios}
\label{subsec:H0_summary}

\begin{table}[]
\centering
\caption{Hubble constant $h$ at 68\% C.L.\ for the
  CMB+DESI+PP\&SH0ES data combination across all scenarios.
  The SH0ES value is $h = 0.7304 \pm 0.0104$.
  Values of $h$ are quoted to three decimal places, consistent
  with their uncertainties.  The last column gives the residual tension
  with SH0ES,
  $N_\sigma=(h_{\rm SH0ES}-h)/\sqrt{\sigma_h^2+\sigma_{\rm SH0ES}^2}$,
  evaluated with the error bar on the side facing the SH0ES value when
  the posterior is asymmetric.}
\label{tab:H0_summary}
\begin{tabular}{lcc}
\multicolumn{1}{c}{\bf Scenario} &
\multicolumn{1}{c}{\bf $h$} &
\multicolumn{1}{c} Tension with SH0ES \\
\hline
BH-I ($\Lambda$CDM)                    & $0.688 \pm 0.003$          & $3.9\sigma$ \\
BH-II ($A$ free)                       & $0.702 \pm 0.007$          & $2.3\sigma$ \\
BTC-I ($m$ free)                       & $0.698^{+0.005}_{-0.006}$  & $2.8\sigma$ \\
BTC-II ($m$, $A$ free)                 & $0.715\pm 0.008$           & $1.2\sigma$ \\
BH ENT-I ($f_B$ free)                  & $0.700\pm 0.006$           & $2.6\sigma$ \\
BH ENT-II ($A$, $f_B$ free)            & $0.702^{+0.006}_{-0.007}$  & $2.4\sigma$ \\
BTC ENT-I ($m$, $f_B$ free)            & $0.703^{+0.006}_{-0.007}$  & $2.3\sigma$ \\
BTC ENT-II ($m$, $A$, $f_B$ free)      & $0.712\pm 0.009$           & $1.3\sigma$ \\
\end{tabular}
\end{table}

Table~\ref{tab:H0_summary}, which collects $h$ for the
CMB+DESI+PP\&SH0ES combination across all eight scenarios, is
the central Hubble-tension result of this work.  Three patterns
emerge.  First, the entanglement correction alone ($f_B$ free)
raises $h$ to $0.6996\pm 0.0059$ (BH ENT-I) and $0.7032$
(BTC ENT-I), reducing the SH0ES tension from ${\sim}3.9\sigma$
to $2.3$--$2.6\sigma$.  Second, models that free $A$
(equivalently $\gamma$) push $h$ to $0.7016$--$0.7148$, reducing
the tension to ${\sim}1.2$--$2.3\sigma$.  Third, the maximum
$h = 0.7148\pm 0.0076$ (BTC-II) still lies ${\sim}1.2\sigma$
below the SH0ES central value: no extended entropy model fully
resolves the Hubble tension among the spatially flat ($k=0$) MHR
scenarios considered here.

A common pattern underlies all of these shifts.  Statistically
significant departures of the extension parameters from their
$\Lambda$CDM values ($A=1$, $m=1$, $f_B=0$) occur only when the
CMB and the SH0ES calibration are combined, reaching
${\sim}2$--$3\sigma$ for BH-II, BTC-II, and BH ENT-I; the
CMB-only chains show no significant departures, and in the
CMB-free combination DESI+PP\&SH0ES+CC $h$ locks onto the SH0ES
value while all extension parameters are consistent with
$\Lambda$CDM within ${\sim}1\sigma$.  The preference for
modified entropy is therefore driven by the CMB--SH0ES
tension---which the additional parameters partially
absorb---rather than by any intrinsic sensitivity of the
late-universe data to modified horizon thermodynamics.  The
Bayesian model comparison below quantifies this conclusion.

\subsection{Bayesian model comparison}
\label{subsec:comparison}

Table~\ref{tab:evidence} collects the Bayesian log-evidence
differences $\Delta\ln\mathcal{Z}$ between the extended scenarios
and the BH-I ($\Lambda$CDM) baseline for all dataset combinations.
\emph{All} extended models are disfavored in every dataset
combination: the values range from $\Delta\ln\mathcal{Z}=-1.0$
(BH-II with CMB+DESI+PP\&SH0ES, weak evidence against) to
$-16.3$ (BTC ENT-II with CMB+DESI+CC, strong evidence against),
and no scenario achieves $\Delta\ln\mathcal{Z}>0$.  The disfavor
is weakest precisely where the $\chi^2$ improvement is largest,
namely in the SH0ES-anchored combinations, but even there it never
turns into a preference.

This verdict is readily understood.  The Bayesian evidence
penalizes unused prior volume through the Occam factor,
$\sim\ln(\Delta\theta/\sigma_\theta)$ per parameter for a prior
of width $\Delta\theta$ and a posterior of width
$\sigma_\theta$; for the scenarios with free $A$ this amounts to
$\ln(50)$ or more per additional parameter, which the
$\Delta\chi^2\approx5$--$16$ improvement from absorbing the
SH0ES-calibrated distance moduli cannot repay.  The
background-viability windows of Paper I~\cite{paper-I} were
deliberately not imposed as priors, so the quoted evidence
values correspond to an agnostic prior choice.  The overall
conclusion is that the extended models fit the SH0ES-anchored
data better than $\Lambda$CDM but are not genuinely preferred:
the improvement reflects the known CMB--SH0ES tension, absorbed
through $A$ or $f_B$ without being resolved.

\subsection{Comparison with previous studies}
\label{sec:lit_comparison}

The MHR deformation exponent $m$ maps directly onto two widely
studied phenomenological entropy modifications of the cosmological
horizon.

\emph{Barrow fractal entropy}~\cite{Barrow:2020tzx}.\ \ In this
framework the horizon acquires a fractal structure parametrized by
$\Delta\in[0,1]$, with the leading-order entropy scaling as
$S_B\propto A^{1+\Delta/2}$.  Comparing with
$S_G\propto r_a^{m+1}\propto A^{(m+1)/2}$ in the $\beta=0$ limit
of Eq.~\eqref{eq:entropy_def} yields
\begin{equation}\label{eq:barrow_map}
  \Delta = m - 1,
\end{equation}
so that $\Delta=0$ ($m=1$) recovers the unmodified
Bekenstein--Hawking entropy.

\emph{Tsallis--Cirto nonextensive entropy}~\cite{Tsallis:1987eu,
Tsallis:2012js}.\ \ Here the entropy takes the form
$S_{TC}\propto A^{\delta}$ with $\delta=1$ the Bekenstein--Hawking
limit; the MHR correspondence (cf.\ Sec.~\ref{sec:res_area}) is
\begin{equation}\label{eq:tsallis_map}
  \delta = \frac{m+1}{2}.
\end{equation}
The two deformation parameters are related by $\delta = 1+\Delta/2$,
so the Tsallis--Cirto deviation $|\delta-1|$ is exactly half the
Barrow deviation $|\Delta|$ at all orders.  The standard
$\Lambda$CDM limit corresponds to $(\Delta,\delta)=(0,1)$.

Table~\ref{tab:barrow_tsallis_comparison} translates the 68\% C.L.\
constraints on $m$ from all four BTC scenarios
(Secs.~\ref{sec:res_area} and~\ref{sec:res_ent}) into $\Delta$ and
$\delta$. No Barrow-domain prior was imposed: BTC-I is
negative for both datasets; BTC-II crosses zero for CMB+DESI but
is positive for Full Data; BTC ENT-I crosses zero for both datasets;
and BTC ENT-II crosses zero for CMB+DESI but is positive for Full
Data.  The table reports the complete, untruncated intervals.  For BTC-II and
BTC ENT-II, the $m$--$\gamma$ degeneracy
(Sec.~\ref{sec:res_area}) broadens the individual $m$ posterior
substantially; the entries for those scenarios reflect the
degeneracy and should not be interpreted as evidence for non-trivial
fractal or non-extensive structure.

\begin{table}[h]
\centering
\caption{ Constraints on the Barrow-mapped exponent $\Delta=m-1$
  and the Tsallis--Cirto nonextensive parameter $\delta=(m+1)/2$ at
  68\% C.L., derived from the four BTC scenarios via
  Eqs.~\eqref{eq:barrow_map} and~\eqref{eq:tsallis_map}.
  No prior $\Delta\in[0,1]$ (or $m\in[1,2]$) was imposed, so the
  entries are complete translations of the MHR posterior; negative
  values lie outside the physical Barrow domain but remain physical
  in the MHR and Tsallis--Cirto frameworks.  The apparent non-zero $\Delta$ in the
  BTC-II and BTC ENT-II with PP\&SH0ES+CC combinations is attributable
  to the $m$--$\gamma$ degeneracy rather than genuine entropy
  modification.}
\label{tab:barrow_tsallis_comparison}
\begin{tabular}{llcc}
\multicolumn{1}{c}{\textbf{Model}} &
\multicolumn{1}{c}{\textbf{Dataset}} &
\multicolumn{1}{c}{$\Delta = m-1$} &
\multicolumn{1}{c}{$\delta - 1 = \Delta/2$} \\
\hline
BTC-I
  & CMB+DESI
  & $(-5.5^{+5.5}_{-1.1})\times10^{-5}$
  & $(-2.8^{+2.8}_{-0.6})\times10^{-5}$ \\
BTC-I
  & Full Data
  & $(-2.1^{+1.3}_{-0.8})\times10^{-4}$
  & $(-1.1^{+0.7}_{-0.4})\times10^{-4}$ \\
\hline
BTC-II
  & CMB+DESI
  & $(9.5^{+5.0}_{-11.0})\times10^{-4}$
  & $(4.8^{+2.5}_{-5.5})\times10^{-4}$ \\
BTC-II
  & Full Data
  & $(2.9\pm1.0)\times10^{-3}$
  & $(1.5\pm0.5)\times10^{-3}$ \\
\hline
BTC ENT-I
  & CMB+DESI
  & $(7.8^{+8.0}_{-15.0})\times10^{-5}$
  & $(3.9^{+4.0}_{-7.5})\times10^{-5}$ \\
BTC ENT-I
  & Full Data
  & $(0.0^{+1.5}_{-1.8})\times10^{-4}$
  & $(0.0^{+0.8}_{-0.9})\times10^{-4}$ \\
\hline
BTC ENT-II
  & CMB+DESI
  & $(4.5^{+8.0}_{-12.0})\times10^{-4}$
  & $(2.3^{+4.0}_{-6.0})\times10^{-4}$ \\
BTC ENT-II
  & Full Data
  & $(2.1\pm1.2)\times10^{-3}$
  & $(1.1\pm0.6)\times10^{-3}$ \\
\end{tabular}
\end{table}

The complete 68\% intervals are listed in
Table~\ref{tab:barrow_tsallis_comparison}.  In particular, the
BTC-I CMB+DESI result has $|\Delta|\lesssim6\times10^{-5}$ and
$|\delta-1|\lesssim3\times10^{-5}$, but its interval is
non-positive---reaching the Barrow boundary $\Delta=0$ only at its
upper edge---and therefore does not lie within the Barrow domain.  The positive Full-Data
intervals for BTC-II and BTC ENT-II are formally within
$\Delta\in[0,1]$, while the intervals that cross zero cannot be
regarded as Barrow-only constraints without rerunning the inference
with $m\in[1,2]$.
The precision originates in the $\sim140$-fold scale-hierarchy
amplification onto $\Omega_{\Lambda0}$
[Sec.~\ref{sec:res_area}, Eq.~\eqref{eq:amplification}].  In
BTC-I the $\gamma=1$ fixing eliminates the $m$--$\gamma$
degeneracy entirely, so this sensitivity reflects the response
of the background data to the freely varying MHR exponent rather than
a projection of the degeneracy ridge.

We now compare with published constraints on $\Delta$ (or the
equivalent MHR exponent $n = m = 1 + \Delta$) from the recent
literature, grouped by physical framework.

\textit{MHR gravity-thermodynamics studies (same framework as BTC).}---
Basilakos et al.~\cite{Basilakos:2025mhr} first constrained the
MHR exponent with late-time data (binned Pantheon SNe\,Ia,
cosmic chronometers, SDSS\,+\,DESI\,DR1 BAO), reporting
$n \simeq 1.1$ ($\Delta \simeq 0.1$) with $\gamma=1$ fixed;
Luciano and Paliathanasis~\cite{Luciano:2025desi} extended this
to DESI\,DR2, Pantheon$+$, and the SH0ES calibration with both
$n$ and $\gamma$ free, finding $n = 0.945\pm0.070$, consistent
with $\Lambda$CDM.  These constraints cannot be compared
directly with our BTC-I bound because the two frameworks
normalize the entropy differently: in the MHR literature
$\gamma$ is dimensionful and normalizes the modification at the
Hubble scale, whereas our dimensionless $\gamma$ normalizes it
at the Planck scale [Eq.~\eqref{eq:entropy_def}], so $n=m=1.1$
implies $A \approx 1.5\times10^{6}$ in our
convention---immediately excluded by any CMB-quality constraint
on $\Omega_{\Lambda0}$.  The $\sim\!10^3$ gap between the two
sets of constraints reflects two compounding effects absent in
the late-time analyses: the sub-percent CMB constraint on
$\Omega_{\Lambda0}$ and the $\sim\!140$-fold Planck-scale
amplification of Eq.~\eqref{eq:amplification}.

Independent bounds from early-Universe physics are consistent
with this picture.  Within the MHR framework,
Luciano~\cite{Luciano:2026pgw} derived $\Delta > -0.116$ from
pulsar-timing-array limits on a primordial gravitational-wave
background, and Luciano and
Saridakis~\cite{Luciano:2025baryogenesis} obtained
$|\Delta| \lesssim 0.02$ from the baryon asymmetry (weaker than
BTC-I by ${\sim}400$, since the Planck-to-horizon amplification
at that scale is only ${\approx}12$).  For generalized entropies
more broadly~\cite{Dabrowski:2025hde}:
$1-\delta<10^{-5}$ from the dark-matter relic
density~\cite{GhoshalLambiase:2021},
$|\delta-1|\approx10^{-3}$ from gravitational
baryogenesis~\cite{LucianoGine:2022},
$\Delta\lesssim10^{-4}$ from inflationary
observables~\cite{Luciano:2023inflation}, and near-standard
exponents from late-time
fits~\cite{AsghariSheykhi:2021,DAgostino:2019}.  These results are
qualitatively compatible with an MHR exponent close to $m=1$, but
they are not directly comparable with the BTC-I posterior, which
imposes no Barrow-domain prior ($m\in[1,2]$).

\textit{Barrow and Tsallis--Cirto holographic dark energy
(different $\Lambda$CDM convention).}---
A distinct class of studies embeds Barrow and Tsallis--Cirto
entropy into a \emph{holographic dark energy} (HDE) scenario,
where the dark-energy density is sourced by the future event
horizon~\cite{Dabrowski:2020barrow,Denkiewicz:2023barrow,Dabrowski:2025hde}.
These analyses consistently find $\Delta > 0.86$ for Barrow--HDE
and $\delta > 1.93$ for Tsallis--Cirto--HDE, with the Bekenstein
value $\Delta = 0$
excluded~\cite{Dabrowski:2020barrow,Denkiewicz:2023barrow} and
Bayesian evidence strongly disfavoring both HDE variants
relative to $\Lambda$CDM~\cite{Dabrowski:2025hde}.  There is no
contradiction with our results: in HDE the density scales as
$\rho_{de}\propto L^{\Delta-2}$ (in their notation), so a
$\Lambda$-like density requires $\Delta\to2$ and large $\Delta$
is built into the viability of the construction---the numerical
values of $\Delta$ are not comparable between the two paradigms.
The BTC-I result is an especially tight constraint on the
MHR exponent---sampled without the Barrow prior ($m\in[1,2]$)---in
the convention where $m=1$ ($\Delta=0$) is the Bekenstein--Hawking
limit; its negative interval should not be presented as a physical
Barrow bound.

\subsection{Expected impact on $S_8$}
\label{sec:S8_outlook}

Many proposed resolutions of the $H_0$ tension worsen the
$S_8$ tension, so it is worth asking how the present framework should
behave.  Collecting Eq.~\eqref{eq:friedmann_flat}, the entire
background modification is a single factor,
\begin{equation}\label{eq:Wfactor}
  H^2\,W(E) = \frac{8\pi G}{3}\rho + \frac{\Lambda c^2}{3},
  \qquad W(E) = A\,E^{1-m} \mp B\,E^{2-\sigma},
\end{equation}
with $W(1)=A\mp B$ and $\Omega_{\Lambda0}=W(1)-(\Omega_{m0}+\Omega_{r0})$
[Eq.~\eqref{eq:OmLambda}]; $\Lambda$CDM is $W\equiv1$.  For the
CMB$+$DESI$+$PP\&SH0ES fits, $W(1)=0.965$--$0.995$ across the seven
extensions.  At recombination $W=0.965$--$0.975$ in all of them, so
that $\Omega_{de}=1-W\simeq2.5$--$3.5\%$ at $z=1100$: at fixed
$\omega_m$ and $\omega_r$ the expansion there is faster than standard
by $1.3$--$1.8\%$, which shortens the sound horizon and is the origin
of the higher $H_0$ inferred at fixed $\theta_s$.  In addition, $W$
itself runs between the two epochs---by up to $3\%$---in every
scenario except BH-II ($m=1$, $\beta=0$), where $W$ is constant and
the modification reduces to the constant rescaling $G_{\rm eff}=G/A$,
$\Lambda_{\rm eff}=\Lambda/A$ of Sec.~\ref{sec:res_area}.  We note
that identifying $W$ with a rescaled coupling more generally,
$G_{\rm eff}=G/W$, is a bookkeeping choice: Eq.~\eqref{eq:ede} makes
the opposite choice, and for $m\neq1$ the factor $W$ depends on $H$
itself, so the background alone does not distinguish them.
Two consequences follow, both of which we state qualitatively.
First, the extensions prefer a slightly lower $\Omega_m$ than
$\Lambda$CDM, which at fixed $\sigma_8$ pushes
$S_8=\sigma_8\sqrt{\Omega_m/0.3}$ downward; the shift stays well
inside the marginalized uncertainty on $S_8$ in every scenario, so it
is not by itself significant.  Second, and more importantly, the
growth response is not fixed by the background at all.  Whatever form
the sub-horizon growth equation eventually takes, the bookkeeping
ambiguity just noted propagates into the effective coupling that
sources it: if that coupling follows the background rescaling
$G_{\rm eff}=G/W$, the growth stays close to $\Lambda$CDM, whereas if
it remains $G$, the growth is suppressed.  The two readings therefore
bracket the outcome between a $\Lambda$CDM-like and a suppressed
$\sigma_8$, and the bracket cannot be narrowed here.
The framework does not yet supply a Poisson equation, and any
quantitative prediction would in addition require the anisotropic
stress, the clustering of the effective dark-energy component, and the
perturbation initial conditions, none of which follow from the
background equations alone.  We therefore attach no number to
$\sigma_8$ or $S_8$ at this stage.
The compressed likelihood cannot settle the question either.  It
constrains only $(R,\ell_a,\omega_b,\omega_c)$ and so carries no
information on $A_s$, $n_s$ or $\tau$; and although $R$ and $\ell_a$
do register the modified expansion above, through $r_s(z_*)$ and
$D_A(z_*)$, they compress it into two numbers and retain nothing of
the acoustic peak morphology, the early ISW effect or the damping
tail, which respond to the same modification.  The fitted
$(\Omega_m,h)$ cannot therefore be assumed reliable at the percent
level for the extensions, and no value of $S_8$ should be inferred
from them.  We conclude only that
nothing in the background analysis indicates that the partial
alleviation of the $H_0$ tension is bought at the cost of a worse
$S_8$ tension; deciding the question requires the perturbation
equations of the model, which have not yet been derived, together with
a full CMB power-spectrum analysis.

\section{Conclusions}
\label{sec:conclusion}
We have carried out a comprehensive observational test of a class
of modified cosmologies derived from a generalized mass-to-horizon
entropy~\eqref{eq:entropy_def}, which is characterized by an
entropy exponent $m$, an MHR coupling parameter $\gamma$ (sampled
via the reparametrized coupling $A$), and a quantum-correction
parameter represented by the rescaled ratio $f_B\equiv B/[(3-m)A]$
(from which the physical exponent $\alpha$ is recovered by
log-Newton inversion).  Using a joint dataset
comprising Pantheon$+$/SH0ES supernovae, DESI~DR2 baryon acoustic
oscillations, Planck~2018 CMB distance priors, and cosmic
chronometers, we constrained eight physical sub-cases via MCMC
sampling and compared models through the Bayesian log-evidence.

The principal conclusions are as follows.

\emph{Tight bound on the entropy exponent.}\ In the scenarios with
fixed MHR coupling parameter ($\gamma=1$),
$|m-1|\lesssim\mathcal{O}(10^{-4})$ at 68\% C.L.\ across all
dataset combinations, relaxing to $\mathcal{O}(10^{-3})$ when
$\gamma$ is free (a projection of the $m$--$\gamma$ degeneracy).
Under $\Delta=m-1$, BTC-I has non-positive 68\% intervals (the
CMB+DESI interval reaching $\Delta=0$ only at its upper edge), BTC-II
and BTC ENT-II have positive Full-Data intervals, and the remaining
listed cases cross zero.  These are complete MHR-posterior
translations because no Barrow prior $m\in[1,2]$ was imposed.
We have demonstrated
that this precision originates from the $\sim\!140$-fold
amplification of $|\delta m|$ onto the effective dark energy
density through the Planck-to-Hubble scale hierarchy
(Eq.~\eqref{eq:amplification}), rather than from any direct
sensitivity of the geometric probes to horizon structure.

\emph{MHR coupling parameter and the Hubble tension.}\ Freeing
$\gamma$ (equivalently, freeing $A$) consistently shifts $H_0$ upward by
$\sim\!1$--$3$~km\,s$^{-1}$\,Mpc$^{-1}$ relative to $\Lambda$CDM,
reducing the CMB--SH0ES tension from $\sim\!4\sigma$ to
$\sim\!1.2$--$2.3\sigma$.  The maximum $h = 0.7148\pm 0.0076$ is
attained by BTC-II (CMB+DESI+PP\&SH0ES), which is about $1.2\sigma$
from the SH0ES determination.  No scenario achieves $H_0$
consistent with SH0ES at the $1\sigma$ level in any CMB-inclusive
chain.  The departure from $A=1$ disappears in the CMB-free
combination, confirming that it is driven by the CMB--SH0ES
tension rather than by genuine evidence for modified horizon
entropy.

\emph{Quantum-correction parameter.}\ In scenarios with $\beta=1$
and $\gamma=1$ (BH ENT-I), sampling via the well-conditioned
ratio $f_B\equiv B/[(3-m)A]$ reveals a nonzero entanglement correction
when Pantheon$+$/SH0ES data are included:
$f_B = 0.0092^{+0.0041}_{-0.0051}$ at $\sim 2\sigma$.
This shifts $h$ from $0.6882$ (CMB+DESI alone) to
$0.6996\pm 0.0059$ (CMB+DESI+PP\&SH0ES), reducing the residual
CMB--SH0ES tension to ${\sim}2.6\sigma$.  This region of
parameter space is accessible only through the well-conditioned
$f_B$ parametrization (Sec.~\ref{sec:reparam}); direct sampling
of the exponentially stretched exponent $\alpha$ mixes poorly and
misses posterior modes.  When $A$ is also freed, the $A$--$f_B$
degeneracy prevents an independent constraint on $f_B$.

\emph{Model comparison.}\ The Bayesian log-evidence disfavors
every extension relative to $\Lambda$CDM in all dataset
combinations, ranging from $\Delta\ln\mathcal{Z}\simeq-1$ (BH-II
with SH0ES-anchored data) to $\simeq-16$ (BTC ENT-II with
CMB+DESI+CC), despite the improved $\chi^2$ fits to the
SH0ES-calibrated distance moduli.  The fit improvement is an
absorption of the CMB--SH0ES tension by the additional
parameters, and it is outweighed by the Occam penalty associated
with their prior volume (Sec.~\ref{subsec:comparison}).  The
present data therefore provide no positive evidence for
departures from Bekenstein--Hawking horizon entropy beyond
$\Lambda$CDM.

\emph{Outlook.}\ The analysis is confined to the background
expansion history.  A complete test of the framework requires the
derivation and constraint of the perturbation equations, including
predictions for the growth rate $f\sigma_8(z)$, the CMB power
spectrum, and the matter power spectrum.  It also requires a full
CMB likelihood analysis (rather than the compressed shift-parameter
approach used here) to validate the CMB constraints for this class
of models.  A comprehensive analysis of the cosmic microwave background,
including a detailed investigation of cosmological perturbations,
is deferred to future work, where the impact of the contained parameters on the growth of structure---and hence on the $S_8$
tension---will determine whether the partial alleviation of the
Hubble tension found here survives at the perturbative level.
The qualitative expectations for $S_8$, and the reasons why its
magnitude cannot be settled \textbf{without the perturbation equations
and} with a compressed CMB likelihood, are discussed in
Sec.~\ref{sec:S8_outlook}.

\clearpage
\bibliographystyle{JHEP}
\bibliography{ref}

@article{paper-I,
    author = "Prasanthan, Pranav and Gohar, Hussain and Salzano, Vincenzo",
    title = "{Modified cosmology from mass-to-horizon relation: Background evolution}",
    eprint = "2607.00133",
    archivePrefix = "arXiv",
    primaryClass = "gr-qc",
    doi = "10.1016/j.physletb.2026.140778",
    journal = "Phys. Lett. B",
    volume = "880",
    pages = "140778",
    year = "2026"
}

@article{Bekenstein:1973ur,
    author = "Bekenstein, Jacob D.",
    title = "{Black holes and entropy}",
    doi = "10.1103/PhysRevD.7.2333",
    journal = "Phys. Rev. D",
    volume = "7",
    pages = "2333--2346",
    year = "1973"
}

@article{Hawking:1974rv,
    author = "Hawking, S. W.",
    title = "{Black hole explosions}",
    doi = "10.1038/248030a0",
    journal = "Nature",
    volume = "248",
    pages = "30--31",
    year = "1974"
}

@article{Cimdiker:2022ics,
    author = "\c{C}imdiker, Ilim and Dabrowski, Mariusz P. and Gohar, Hussain",
    title = "{Equilibrium temperature for black holes with nonextensive entropy}",
    eprint = "2208.04473",
    archivePrefix = "arXiv",
    primaryClass = "gr-qc",
    doi = "10.1140/epjc/s10052-023-11317-0",
    journal = "Eur. Phys. J. C",
    volume = "83",
    number = "2",
    pages = "169",
    year = "2023"
}

@article{Gohar:2023hnb,
    author = "Gohar, Hussain and Salzano, Vincenzo",
    title = "{On the foundations of entropic cosmologies: Inconsistencies, possible solutions and dead end signs}",
    eprint = "2307.01768",
    archivePrefix = "arXiv",
    primaryClass = "gr-qc",
    doi = "10.1016/j.physletb.2024.138781",
    journal = "Phys. Lett. B",
    volume = "855",
    pages = "138781",
    year = "2024"
}

@article{Tsallis:1987eu,
    author = "Tsallis, Constantino",
    title = "{Possible Generalization of Boltzmann-Gibbs Statistics}",
    reportNumber = "CBPF-NF-062-87",
    doi = "10.1007/BF01016429",
    journal = "J. Statist. Phys.",
    volume = "52",
    pages = "479--487",
    year = "1988"
}

@article{reny1,
    author = "Renyi, A.",
    title = "{On the dimension and entropy of probability distributions}",
    doi = "https://doi.org/10.1007/BF02063299",
    journal = "Acta Mathematica Academiae Scientiarum Hungarica",
    volume = "10",
    pages = "193--215",
    year = "1959"
}

@article{Gong:2007md,
    author = "Gong, Yungui and Wang, Anzhong",
    title = "{The Friedmann equations and thermodynamics of apparent horizons}",
    eprint = "0704.0793",
    archivePrefix = "arXiv",
    primaryClass = "hep-th",
    doi = "10.1103/PhysRevLett.99.211301",
    journal = "Phys. Rev. Lett.",
    volume = "99",
    pages = "211301",
    year = "2007"
}

@article{Padmanabhan:2003pk,
    author = "Padmanabhan, T.",
    title = "{Gravitational entropy of static space-times and microscopic density of states}",
    eprint = "gr-qc/0308070",
    archivePrefix = "arXiv",
    doi = "10.1088/0264-9381/21/18/013",
    journal = "Class. Quant. Grav.",
    volume = "21",
    pages = "4485--4494",
    year = "2004"
}

@article{Jacobson:1995ab,
    author = "Jacobson, Ted",
    title = "{Thermodynamics of space-time: The Einstein equation of state}",
    eprint = "gr-qc/9504004",
    archivePrefix = "arXiv",
    reportNumber = "UMDGR-95-114",
    doi = "10.1103/PhysRevLett.75.1260",
    journal = "Phys. Rev. Lett.",
    volume = "75",
    pages = "1260--1263",
    year = "1995"
}

@article{Cai:2005ra,
    author = "Cai, Rong-Gen and Kim, Sang Pyo",
    title = "{First law of thermodynamics and Friedmann equations of Friedmann-Robertson-Walker universe}",
    eprint = "hep-th/0501055",
    archivePrefix = "arXiv",
    doi = "10.1088/1126-6708/2005/02/050",
    journal = "JHEP",
    volume = "02",
    pages = "050",
    year = "2005"
}

@article{Medved:2004yu,
    author = "Medved, A. J. M. and Vagenas, Elias C.",
    title = "{When conceptual worlds collide: The GUP and the BH entropy}",
    eprint = "hep-th/0411022",
    archivePrefix = "arXiv",
    reportNumber = "UB-ECM-PF-04-33",
    doi = "10.1103/PhysRevD.70.124021",
    journal = "Phys. Rev. D",
    volume = "70",
    pages = "124021",
    year = "2004"
}

@article{Meissner:2004ju,
    author = "Meissner, Krzysztof A.",
    title = "{Black hole entropy in loop quantum gravity}",
    eprint = "gr-qc/0407052",
    archivePrefix = "arXiv",
    doi = "10.1088/0264-9381/21/22/015",
    journal = "Class. Quant. Grav.",
    volume = "21",
    pages = "5245--5252",
    year = "2004"
}

@article{Rovelli:1996dv,
    author = "Rovelli, Carlo",
    title = "{Black hole entropy from loop quantum gravity}",
    eprint = "gr-qc/9603063",
    archivePrefix = "arXiv",
    doi = "10.1103/PhysRevLett.77.3288",
    journal = "Phys. Rev. Lett.",
    volume = "77",
    pages = "3288--3291",
    year = "1996"
}

@article{Das:2007mj,
    author = "Das, Saurya and Shankaranarayanan, S. and Sur, Sourav",
    title = "{Power-law corrections to entanglement entropy of black holes}",
    eprint = "0705.2070",
    archivePrefix = "arXiv",
    primaryClass = "gr-qc",
    doi = "10.1103/PhysRevD.77.064013",
    journal = "Phys. Rev. D",
    volume = "77",
    pages = "064013",
    year = "2008"
}

@article{Verlinde:2010hp,
    author = "Verlinde, Erik P.",
    title = "{On the Origin of Gravity and the Laws of Newton}",
    eprint = "1001.0785",
    archivePrefix = "arXiv",
    primaryClass = "hep-th",
    doi = "10.1007/JHEP04(2011)029",
    journal = "JHEP",
    volume = "04",
    pages = "029",
    year = "2011"
}

@article{Gohar:2025yfx,
    author = "Gohar, Hussain",
    title = "{Mass-to-Horizon relation and entropy beyond the Bekenstein-Hawking limit}",
    eprint = "2510.07587",
    archivePrefix = "arXiv",
    primaryClass = "gr-qc",
    doi = "10.1016/j.physletb.2026.140148",
    journal = "Phys. Lett. B",
    volume = "873",
    pages = "140148",
    year = "2026"
}

@article{Brout:2022vxf,
    author = "Brout, Dillon and others",
    title = "{The Pantheon+ Analysis: Cosmological Constraints}",
    eprint = "2202.04077",
    archivePrefix = "arXiv",
    primaryClass = "astro-ph.CO",
    doi = "10.3847/1538-4357/ac8e04",
    journal = "Astrophys. J.",
    volume = "938",
    number = "2",
    pages = "110",
    year = "2022"
}

@article{Planck:2018vyg,
    author = "Aghanim, N. and others",
    collaboration = "Planck",
    title = "{Planck 2018 results. VI. Cosmological parameters}",
    eprint = "1807.06209",
    archivePrefix = "arXiv",
    primaryClass = "astro-ph.CO",
    doi = "10.1051/0004-6361/201833910",
    journal = "Astron. Astrophys.",
    volume = "641",
    pages = "A6",
    year = "2020",
    note = "[Erratum: Astron.Astrophys. 652, C4 (2021)]"
}

@article{Wang:2007mza,
    author = "Wang, Yun and Mukherjee, Pia",
    title = "{Observational Constraints on Dark Energy and Cosmic Curvature}",
    eprint = "astro-ph/0703780",
    archivePrefix = "arXiv",
    doi = "10.1103/PhysRevD.76.103533",
    journal = "Phys. Rev. D",
    volume = "76",
    pages = "103533",
    year = "2007"
}

@article{Hogg:1999ad,
    author = "Hogg, David W.",
    title = "{Distance measures in cosmology}",
    eprint = "astro-ph/9905116",
    archivePrefix = "arXiv",
    month = "5",
    year = "1999"
}

@article{Moresco:2010wh,
    author = "Moresco, Michele and Jimenez, Raul and Cimatti, Andrea and Pozzetti, Lucia",
    title = "{Constraining the expansion rate of the Universe using low-redshift ellipticals as cosmic chronometers}",
    eprint = "1010.0831",
    archivePrefix = "arXiv",
    primaryClass = "astro-ph.CO",
    doi = "10.1088/1475-7516/2011/03/045",
    journal = "JCAP",
    volume = "03",
    pages = "045",
    year = "2011"
}

@article{Jimenez:2001gg,
    author = "Jimenez, Raul and Loeb, Abraham",
    title = "{Constraining cosmological parameters based on relative galaxy ages}",
    eprint = "astro-ph/0106145",
    archivePrefix = "arXiv",
    doi = "10.1086/340549",
    journal = "Astrophys. J.",
    volume = "573",
    pages = "37--42",
    year = "2002"
}

@article{Moresco:2018xdr,
    author = "Moresco, Michele and Jimenez, Raul and Verde, Licia and Pozzetti, Lucia and Cimatti, Andrea and Citro, Annalisa",
    title = "{Setting the Stage for Cosmic Chronometers. I. Assessing the Impact of Young Stellar Populations on Hubble Parameter Measurements}",
    eprint = "1804.05864",
    archivePrefix = "arXiv",
    primaryClass = "astro-ph.CO",
    doi = "10.3847/1538-4357/aae829",
    journal = "Astrophys. J.",
    volume = "868",
    number = "2",
    pages = "84",
    year = "2018"
}

@article{Moresco:2020fbm,
    author = "Moresco, Michele and Jimenez, Raul and Verde, Licia and Cimatti, Andrea and Pozzetti, Lucia",
    title = "{Setting the Stage for Cosmic Chronometers. II. Impact of Stellar Population Synthesis Models Systematics and Full Covariance Matrix}",
    eprint = "2003.07362",
    archivePrefix = "arXiv",
    primaryClass = "astro-ph.GA",
    doi = "10.3847/1538-4357/ab9eb0",
    journal = "Astrophys. J.",
    volume = "898",
    number = "1",
    pages = "82",
    year = "2020"
}

@article{Moresco:2022phi,
    author = "Moresco, Michele and others",
    title = "{Unveiling the Universe with emerging cosmological probes}",
    eprint = "2201.07241",
    archivePrefix = "arXiv",
    primaryClass = "astro-ph.CO",
    doi = "10.1007/s41114-022-00040-z",
    journal = "Living Rev. Rel.",
    volume = "25",
    number = "1",
    pages = "6",
    year = "2022"
}

@article{Jiao:2022aep,
    author = "Jiao, Kang and Borghi, Nicola and Moresco, Michele and Zhang, Tong-Jie",
    title = "{New Observational H(z) Data from Full-spectrum Fitting of Cosmic Chronometers in the LEGA-C Survey}",
    eprint = "2205.05701",
    archivePrefix = "arXiv",
    primaryClass = "astro-ph.CO",
    doi = "10.3847/1538-4365/acbc77",
    journal = "Astrophys. J. Suppl.",
    volume = "265",
    number = "2",
    pages = "48",
    year = "2023"
}

@article{Gohar:2023lta,
    author = "Gohar, Hussain and Salzano, Vincenzo",
    title = "{Generalized mass-to-horizon relation: A new global approach to entropic cosmologies and its connection to \ensuremath{\Lambda}CDM}",
    eprint = "2307.06239",
    archivePrefix = "arXiv",
    primaryClass = "gr-qc",
    doi = "10.1103/PhysRevD.109.084075",
    journal = "Phys. Rev. D",
    volume = "109",
    number = "8",
    pages = "084075",
    year = "2024"
}

@article{Zhai:2019nad,
    author = "Zhai, Zhongxu and Park, Chan-Gyung and Wang, Yun and Ratra, Bharat",
    title = "{CMB distance priors revisited: effects of dark energy dynamics, spatial curvature, primordial power spectrum, and neutrino parameters}",
    eprint = "1912.04921",
    archivePrefix = "arXiv",
    primaryClass = "astro-ph.CO",
    doi = "10.1088/1475-7516/2020/07/009",
    journal = "JCAP",
    volume = "07",
    pages = "009",
    year = "2020"
}

@article{emcee,
   author = {{Foreman-Mackey}, D. and {Hogg}, D.~W. and {Lang}, D. and {Goodman}, J.},
    title = {emcee: The MCMC Hammer},
  journal = {PASP},
     year = 2013,
   volume = 125,
    pages = {306-312},
   eprint = {1202.3665},
      doi = {10.1086/670067}
}

@article{Tsallis:2012js,
    author = "Tsallis, Constantino and Cirto, Leonardo J. L.",
    title = "{Black hole thermodynamical entropy}",
    eprint = "1202.2154",
    archivePrefix = "arXiv",
    primaryClass = "cond-mat.stat-mech",
    doi = "10.1140/epjc/s10052-013-2487-6",
    journal = "Eur. Phys. J. C",
    volume = "73",
    pages = "2487",
    year = "2013"
}

@article{Barrow:2020tzx,
    author = "Barrow, John D.",
    title = "{The Area of a Rough Black Hole}",
    eprint = "2004.09444",
    archivePrefix = "arXiv",
    primaryClass = "gr-qc",
    doi = "10.1016/j.physletb.2020.135643",
    journal = "Phys. Lett. B",
    volume = "808",
    pages = "135643",
    year = "2020"
}

@article{Aizpuru:2021vhd,
    author = "Aizpuru, Andoni and Arjona, Rub\'en and Nesseris, Savvas",
    title = "{Machine learning improved fits of the sound horizon at the baryon drag epoch}",
    eprint = "2106.00428",
    archivePrefix = "arXiv",
    primaryClass = "astro-ph.CO",
    reportNumber = "IFT-UAM/CSIC-21-67",
    doi = "10.1103/PhysRevD.104.043521",
    journal = "Phys. Rev. D",
    volume = "104",
    number = "4",
    pages = "043521",
    year = "2021"
}

@article{DESI:2025zgx,
    author = "Abdul Karim, M. and others",
    collaboration = "DESI",
    title = "{DESI DR2 results. II. Measurements of baryon acoustic oscillations and cosmological constraints}",
    eprint = "2503.14738",
    archivePrefix = "arXiv",
    primaryClass = "astro-ph.CO",
    reportNumber = "FERMILAB-PUB-25-0169-PPD",
    doi = "10.1103/tr6y-kpc6",
    journal = "Phys. Rev. D",
    volume = "112",
    number = "8",
    pages = "083515",
    year = "2025"
}

@article{Carr:2021lcj,
    author = "Carr, Anthony and Davis, Tamara M. and Scolnic, Dan and Scolnic, Daniel and Said, Khaled and Brout, Dillon and Peterson, Erik R. and Kessler, Richard",
    title = "{The Pantheon+ analysis: Improving the redshifts and peculiar velocities of Type Ia supernovae used in cosmological analyses}",
    eprint = "2112.01471",
    archivePrefix = "arXiv",
    primaryClass = "astro-ph.CO",
    doi = "10.1017/pasa.2022.41",
    journal = "Publ. Astron. Soc. Austral.",
    volume = "39",
    pages = "e046",
    year = "2022"
}

@article{Gohar:2026hiy,
    author = "Gohar, Hussain",
    title = "{Scalar$-$Tensor Gravity as a Probe of Generalized Black Hole Entropy}",
    eprint = "2605.18551",
    archivePrefix = "arXiv",
    primaryClass = "gr-qc",
    month = "5",
    year = "2026"
}

@article{Denkiewicz:2025txx,
    author = "Denkiewicz, Tomasz and Gohar, Hussain",
    title = "{Mass-to-horizon entropic cosmology: A unified thermodynamic pathway to cosmic acceleration}",
    eprint = "2512.22103",
    archivePrefix = "arXiv",
    primaryClass = "gr-qc",
    doi = "10.1103/zqkb-bqth",
    journal = "Phys. Rev. D",
    volume = "113",
    number = "6",
    pages = "063564",
    year = "2026"
}

@article{Foreman-Mackey:2012any,
    author = "Foreman-Mackey, Daniel and Hogg, David W. and Lang, Dustin and Goodman, Jonathan",
    title = "{emcee: The MCMC Hammer}",
    eprint = "1202.3665",
    archivePrefix = "arXiv",
    primaryClass = "astro-ph.IM",
    doi = "10.1086/670067",
    journal = "Publ. Astron. Soc. Pac.",
    volume = "125",
    pages = "306--312",
    year = "2013"
}

@article{Heavens:2017afc,
    author = "Heavens, Alan and Fantaye, Yabebal and Mootoovaloo, Arrykrishna and Eggers, Hans and Hosenie, Zafiirah and Kroon, Steve and Sellentin, Elena",
    title = "{Marginal Likelihoods from Monte Carlo Markov Chains}",
    eprint = "1704.03472",
    archivePrefix = "arXiv",
    primaryClass = "stat.CO",
    month = "4",
    year = "2017"
}

@article{Basilakos:2025mhr,
  author        = {Basilakos, Spyros and Lymperis, Alexis and
                   Petronikolou, Maria and Saridakis, Emmanuel N.},
  title         = {Modified cosmology through spacetime thermodynamics
                   and generalized mass-to-horizon entropy},
  journal       = {arXiv},
  year          = {2025},
  eprint        = {2503.24355},
  archivePrefix = {arXiv},
  primaryClass  = {gr-qc},
}

@article{Luciano:2025desi,
  author        = {Luciano, Giuseppe Gaetano and Paliathanasis, Andronikos},
  title         = {Modified cosmology through generalized
                   mass-to-horizon entropy: observational constraints
                   from {DESI DR2 BAO} data},
  journal       = {Phys. Lett. B},
  volume        = {870},
  pages         = {139954},
  year          = {2025},
  doi           = {10.1016/j.physletb.2025.139954},
  eprint        = {2508.13260},
  archivePrefix = {arXiv},
  primaryClass  = {gr-qc},
}

@article{Luciano:2026pgw,
  author        = {Luciano, Giuseppe Gaetano},
  title         = {Modified cosmology through generalized
                   mass-to-horizon ratio: implications for structure
                   growth and primordial gravitational waves},
  journal       = {JHEAp},
  volume        = {50},
  pages         = {100487},
  year          = {2026},
  doi           = {10.1016/j.jheap.2025.100487},
  eprint        = {2510.00673},
  archivePrefix = {arXiv},
  primaryClass  = {gr-qc},
}

@article{Luciano:2025baryogenesis,
  author        = {Luciano, Giuseppe Gaetano and Saridakis, Emmanuel N.},
  title         = {Baryogenesis constraints on generalized
                   mass-to-horizon entropy},
  journal       = {arXiv},
  year          = {2025},
  eprint        = {2511.01693},
  archivePrefix = {arXiv},
  primaryClass  = {gr-qc},
}

@article{Denkiewicz:2023barrow,
  author        = {Denkiewicz, Tomasz and Salzano, Vincenzo and
                   D\k{a}browski, Mariusz P.},
  title         = {Barrow nearly-extensive {G}ibbs-like entropy
                   favoured by the full dynamical and geometrical
                   data set in cosmology},
  journal       = {Phys. Rev. D},
  volume        = {108},
  pages         = {103533},
  year          = {2023},
  doi           = {10.1103/PhysRevD.108.103533},
  eprint        = {2303.11680},
  archivePrefix = {arXiv},
  primaryClass  = {astro-ph.CO},
}

@article{Dabrowski:2025hde,
  author        = {{\c{C}}imdiker, {\.I}lim and D\k{a}browski, Mariusz P.
                   and Salzano, Vincenzo},
  title         = {Generalized Nonextensive Entropy Holographic Dark
                   Energy Models Verified by Cosmological Data},
  journal       = {arXiv},
  year          = {2025},
  eprint        = {2503.18230},
  archivePrefix = {arXiv},
  primaryClass  = {astro-ph.CO},
}

@article{Dabrowski:2020barrow,
  author        = {D\k{a}browski, Mariusz P. and Salzano, Vincenzo},
  title         = {Geometrical observational bounds on a fractal horizon
                   holographic dark energy},
  journal       = {Phys. Rev. D},
  volume        = {102},
  pages         = {064047},
  year          = {2020},
  doi           = {10.1103/PhysRevD.102.064047},
  eprint        = {2009.08306},
  archivePrefix = {arXiv},
  primaryClass  = {astro-ph.CO},
}

@article{LucianoGine:2022,
  author        = {Luciano, Giuseppe Gaetano and Gin\'{e}, Jaume},
  title         = {Baryogenesis in non-extensive {Tsallis} cosmology},
  journal       = {Phys. Lett. B},
  volume        = {833},
  pages         = {137352},
  year          = {2022},
  doi           = {10.1016/j.physletb.2022.137352},
  eprint        = {2204.02723},
  archivePrefix = {arXiv},
  primaryClass  = {gr-qc},
}

@article{GhoshalLambiase:2021,
  author        = {Ghoshal, Anish and Lambiase, Gaetano},
  title         = {Constraints on {Tsallis} cosmology from {Big Bang}
                   nucleosynthesis and dark matter freeze-out},
  journal       = {arXiv},
  year          = {2021},
  eprint        = {2104.11296},
  archivePrefix = {arXiv},
  primaryClass  = {astro-ph.CO},
}

@article{Luciano:2023inflation,
  author        = {Luciano, Giuseppe Gaetano},
  title         = {Constraining {Barrow} entropy-based cosmology
                   with power-law inflation},
  journal       = {Eur. Phys. J. C},
  volume        = {83},
  pages         = {329},
  year          = {2023},
  doi           = {10.1140/epjc/s10052-023-11499-7},
  eprint        = {2301.12509},
  archivePrefix = {arXiv},
  primaryClass  = {gr-qc},
}

@article{AsghariSheykhi:2021,
  author        = {Asghari, Mahnaz and Sheykhi, Ahmad},
  title         = {Observational constraints on {Tsallis} modified gravity},
  journal       = {Mon. Not. Roy. Astron. Soc.},
  volume        = {508},
  pages         = {2855},
  year          = {2021},
  doi           = {10.1093/mnras/stab2671},
  eprint        = {2106.15551},
  archivePrefix = {arXiv},
  primaryClass  = {gr-qc},
}

@article{DAgostino:2019,
  author        = {D'Agostino, Rocco},
  title         = {Holographic dark energy from nonadditive entropy:
                   cosmological perturbations and observational constraints},
  journal       = {Phys. Rev. D},
  volume        = {99},
  pages         = {103524},
  year          = {2019},
  doi           = {10.1103/PhysRevD.99.103524},
  eprint        = {1903.03836},
  archivePrefix = {arXiv},
  primaryClass  = {gr-qc},
}

\end{document}